\documentclass[10pt,aps,prd,twocolumn,notitlepage,bibnotes,longbibliography,
floatfix,showpacs,citeautoscript,superscriptaddress]{revtex4}

\usepackage[english]{babel}		

\usepackage{amsmath}		
\usepackage{amsfonts}		
\usepackage{amssymb}		
\usepackage{bbm}			
\usepackage{physics}		

\usepackage{graphicx}		
\usepackage{graphics}		
\usepackage{array}		

\usepackage{dsfont}
\usepackage{ifthen}
\usepackage{ulem}
\usepackage{float}
\usepackage{bbold}
\usepackage{lipsum}
\usepackage{nccmath}

\usepackage[pdfstartview=FitV, pdfusetitle,
			bookmarks=true,bookmarksnumbered=false,bookmarksopen=false,
			breaklinks=true,pdfborder={0 0 0},backref=false,colorlinks=true, 
			linkcolor=blue, citecolor=blue, urlcolor=blue]{hyperref}

\newcommand{\bkfa}{Ba$_{1-x}$K$_x$Fe$_2$As$_2$}

\newcommand{\RNum}[1]{\uppercase\expandafter{\romannumeral #1\relax}}
\newcommand{\RNeq}[1]{\textup{\uppercase\expandafter{\romannumeral#1}}}

\makeatletter
\newcommand*{\Equation}{\@ifstar\sEquation\oEquation}
\newcommand{\sEquation}[1]{\begin{equation*}#1\end{equation*}}
\newcommand{\oEquation}[2]{  \begin{equation}\label{#1}#2\end{equation} }
\makeatother
\newcommand{\Align}[2]{\begin{align}\label{#1}#2\end{align}}
\newcommand{\SubAlign}[2]{\begin{subequations}\label{#1}\begin{align}#2\end{align}\end{subequations}}
\newcommand{\mbf}{\mathbf}
 
\newcommand{\bs}{\boldsymbol}
\newcommand{\Figref}[1]{Fig.~\ref{#1}}
\newcommand{\Eqref}[1]{\eqref{#1}}
\newcommand{\groupZ}[1]{\mathbb{Z}_{#1}} 			
\newcommand{\groupU}[1]{\mathrm{U}(#1)}  			
\newcommand{\Exp}[1]{\text{e}^{#1}}

\renewcommand\Im{\mathrm{Im}}
\newcommand{\Grad}{{\bs\nabla}}

\newcommand{\Curl}{{\bs\nabla}\times}
\newcommand{\F}{\mathcal{F}}
\newcommand{\A}{{\bs A}}
\newcommand{\B}{{\bs B}}
\newcommand{\D}{{\bs D}}

\newcommand{\Tc}[1]{{T_{\rm c}^{#1}}}   			
\newcommand{\Hc}[1]{H_{c#1}}   			
\newcommand{\sis}{{s\!+\!is}}
\newcommand{\Ztwo}{{\groupZ{2}}}
\newcommand{\TcZtwo}{{\Tc{\groupZ{2}}}}

\usepackage{color}
\usepackage{xcolor}
\usepackage{comment} 

\begin{document}

\title{Evidence of pseudogap and absence of spin magnetism in the time-reversal-symmetry-breaking state of \texorpdfstring{\bkfa}{Ba1-xKxFe2As2} }

\author{Florian B\"artl}
\email{f.baertl@hzdr.de}
\affiliation{Dresden High Magnetic Field Laboratory (HLD-EMFL) and Würzburg-Dresden Cluster of Excellence ct.qmat, Helmholtz-Zentrum Dresden-Rossendorf, 01328 Dresden, Germany}
\affiliation{Institute for Solid State and Materials Physics, Technische Universit\"at Dresden, 01062 Dresden, Germany}
\author{Nadia Stegani}
\affiliation{University of Genoa, Via Dodecaneso 33, 16146 Genoa, Italy}
\affiliation{Consiglio Nazionale delle Ricerche (CNR)-SPIN, Corso Perrone 24, 16152 Genova, Italy}
\author{Federico Caglieris}
\affiliation{Consiglio Nazionale delle Ricerche (CNR)-SPIN, Corso Perrone 24, 16152 Genova, Italy}
\author{Ilya Shipulin}
\affiliation{Institute for Solid State and Materials Physics, Technische Universit\"at Dresden, 01062 Dresden, Germany}
\author{Yongwei Li}
\affiliation{Tsung-Dao Lee Institute \& School of Physics and Astronomy, Shanghai Jiao Tong University, Shanghai 201210, China}
\author{Quanxin Hu}
\affiliation{Tsung-Dao Lee Institute \& School of Physics and Astronomy, Shanghai Jiao Tong University, Shanghai 201210, China}
\author{Yu Zheng}
\affiliation{Tsung-Dao Lee Institute \& School of Physics and Astronomy, Shanghai Jiao Tong University, Shanghai 201210, China}
\author{Chi-Ming Yim}
\affiliation{Tsung-Dao Lee Institute \& School of Physics and Astronomy, Shanghai Jiao Tong University, Shanghai 201210, China}
\author{Sven Luther}
\affiliation{Dresden High Magnetic Field Laboratory (HLD-EMFL) and Würzburg-Dresden Cluster of Excellence ct.qmat, Helmholtz-Zentrum Dresden-Rossendorf, 01328 Dresden, Germany}
\author{Jochen Wosnitza}
\affiliation{Dresden High Magnetic Field Laboratory (HLD-EMFL) and Würzburg-Dresden Cluster of Excellence ct.qmat, Helmholtz-Zentrum Dresden-Rossendorf, 01328 Dresden, Germany}
\affiliation{Institute for Solid State and Materials Physics, Technische Universit\"at Dresden, 01062 Dresden, Germany}
\author{Rajib Sarkar}
\affiliation{Institute for Solid State and Materials Physics, Technische Universit\"at Dresden, 01062 Dresden, Germany}
\author{Hans-Henning Klauss}
\affiliation{Institute for Solid State and Materials Physics, Technische Universit\"at Dresden, 01062 Dresden, Germany}
\author{Julien Garaud}
\affiliation{Institut Denis Poisson UMR CNRS 7013, Universit\'e de Tours, 37200 Tours, France}
\author{Egor Babaev}
\email{babaev@kth.se}
\affiliation{Department of Physics, KTH Royal Institute of Technology, SE-106 91 Stockholm, Sweden}
\affiliation{Wallenberg Initiative Materials Science for Sustainability, Department of Physics, KTH Royal Institute of Technology, SE-106 91 Stockholm, Sweden}
\author{Hannes K\"uhne}
\affiliation{Dresden High Magnetic Field Laboratory (HLD-EMFL) and Würzburg-Dresden Cluster of Excellence ct.qmat, Helmholtz-Zentrum Dresden-Rossendorf, 01328 Dresden, Germany}

\author{Vadim Grinenko}
\email{vadim.grinenko@sjtu.edu.cn}
\affiliation{Tsung-Dao Lee Institute \& School of Physics and Astronomy, Shanghai Jiao Tong University, Shanghai 201210, China}

\begin{abstract}
Muon-spin-rotation ($\mu SR$) experiments and the observation of a spontaneous Nernst effect indicate time-reversal symmetry breaking (BTRS) at $\TcZtwo$ above the superconducting transition temperature $\Tc{}$ in \bkfa{}, with $x\approx0.8$. 
Further studies have pointed out that BTRS is caused by the formation of a new state of matter associated with the condensation of pairs of electron pairs. 
Despite exhibiting multiple unconventional effects that warrant further investigation, the electronic spectral properties of this electron quadrupling state remain largely unexplored.
 Here, we present detailed $^{75}$As nuclear magnetic resonance (NMR) measurements of \bkfa{}, with $x = 0.77$, which has $\TcZtwo > \Tc{}$ according to measurements of the spontaneous Nernst effect. The NMR data obtained in this work provide the first direct electronic spectral characteristics of the electron quadrupling state by indicating that it evolves from a pseudogap that sets in at $T^*$ well above $\TcZtwo$. This pseudogap behavior is consistent with $\mu$SR Knight-shift, specific-heat, and transport data indicating the formation of a bound state of electrons. According to a theory of electron quadrupling condensates, such bound-state formations should precede the onset of BTRS correlations between pairs of electron pairs. 
The second important insight from NMR data is the absence of spin-related magnetism. The temperature dependence of the spin-lattice relaxation rate $1/T_1T$ and the evolution of the NMR linewidth prove the absence of a magnetic transition at $\TcZtwo$ and rule out even a proximity to some magnetic instability. 
This indicates that the spontaneous magnetic fields detected in this compound are not caused by spin magnetism but are associated with persistent real-space currents.

 \end{abstract}
 
\maketitle

\section{Introduction}

The experiments on \bkfa{} in a narrow range of a specific, recently termed ``magic" doping level of $x\approx0.8$, revealed a number of unprecedented and unconventional effects. This includes the observation of vortices that carry a fraction of the magnetic flux quantum in the superconducting state \cite{Iguchi2023,zhou2024observation}. Importantly, the fractionalization of vortex cores was demonstrated on KFe$_2$As$_2$ surface when the surface yields a doping level similar to magic doping~\cite{zheng2024direct}. Particular unusual observations were made in the adjacent non-superconducting state of \bkfa, that were interpreted as the formation of a new state of matter, namely, a four-electron condensate that breaks time-reversal symmetry \cite{Grinenko2021state,shipulin2023calorimetric,halcrow2024probing}. This motivates further experimental investigations of this material.

At first glance, the presence of conventional magnetic phases in \bkfa{} at the magic doping level appears to be a plausible possibility, considering the system's rich phase diagram. At low potassium (K) doping level, this material is an antiferromagnetic metal with a long-range spin density wave (SDW) state \cite{Aczel2008}. With increasing K doping, the SDW phase vanishes. Superconductivity coexists with magnetism above $x \approx$ 0.1, and magnetism is completely suppressed at $x \approx$ 0.3 \cite{avci2012}. The superconducting critical temperature ($\Tc{}$) reaches its maximum at $x \approx$ 0.4 and monotonically decreases with further increase in doping. The potassium substitution introduces one hole per atom, resulting in a gradual shift of the Fermi level, which leads to the disappearance of the electron Fermi pockets at the M point of the Brillouin zone at $x \approx$ 0.7 (Lifshitz transition) and the appearance of hole pockets (for more details see \cite{Grinenko2020superconductivity}). Initially, it was believed that the phase diagram has two separate superconducting domes for \bkfa{} at low and high doping levels \cite{Chubukov2012}. Later, it was found that the superconducting dome continues up to the complete substitution of Ba by K. However, the end member KFe$_2$As$_2$ has a very high Sommerfeld coefficient of the electronic specific heat and may be close to another magnetic quantum critical point \cite{Hardy2016, Dong2010, Wu2016}. Therefore, it was speculated that this dome might be in proximity to another type of magnetism \cite{Drechsler2018}.
 
Close to the Lifshitz transition, the superconducting state is unusual. A narrow superconductivity dome with broken time-reversal symmetry (BTRS) appears around $x \approx$ 0.8 in muon-spin-rotation/relaxation ($\mu$SR) measurements \cite{Grinenko2017bkfa,Grinenko2020superconductivity}. The experimental results are consistent with an $s+is$ superconducting state, but other similar states, such as $s+id$, are not ruled out. A $s+is$ superconductor is a multiband superconductor that has a non-trivial difference between the phases of the gaps in different bands. This occurs because of frustration of the interband Josephson interactions. When time-reversal symmetry is broken, spontaneous magnetic fields appear near certain kinds of defects and domain walls \cite{garaud2014domain,Maiti2013,benfenati2020magnetic}, consistent with observations by $\mu$SR 
\cite{Grinenko2020superconductivity}. The frustration affects phase fluctuations, which are responsible for a significantly suppressed $\Tc{}$ in \bkfa, compared to the temperature $T^*$, at which incoherent Cooper pairs are formed. 

It was predicted that the fluctuations 
can change the sequence of the phase transitions, resulting in the formation of the BTRS state above $\Tc{}$~\cite{Bojesen2013time,bojesen2014phase,Maccari2022effects}. These predictions are supported by recent experimental data \cite{Grinenko2021state}. $\mu$SR and spontaneous-Nernst-effect measurements for samples with doping levels $x$ between 0.75 and 0.82 indicate that time-reversal symmetry is already broken at $\TcZtwo > \Tc{}$. Further, indication for superconducting fluctuations below $T^* \sim 2\Tc{}$ was found in electric and thermoelectric transport measurements \cite{Grinenko2021state,shipulin2023calorimetric}. The phase transition at $\TcZtwo$ was recently confirmed by the observation of an anomaly in the specific heat above $\Tc{}$ \cite{shipulin2023calorimetric}. Overall, the data give evidence for a quartic metal state, which is different in symmetry from the normal metallic and superconducting state. However, the magnetic nature of the fluctuations cannot be excluded based on existing data, which calls for further studies with probes specifically sensitive to dynamic magnetism.

The state above the superconducting phase that spontaneously breaks time-reversal symmetry is described by a four-electron order parameter, hence termed the quartic state. It is a correlated state, in which two electrons in one of the bands allow certain dissipationless anticorrelation with two electrons in another band. Very recently, the fundamental possibility for this mechanism was demonstrated directly by the realization of the counterflow superfluidity in a two-component Mott insulator using cold atoms \cite{Zheng2025}.
Phenomenologically, the quartic state results in magnetic effects, which differ from the usual magnetic state related to spin order. Time-reversal symmetry, in this case, is broken by spontaneous Josephson interband currents, resulting in persistent real-space loop-currents near defects~\cite{Garaud2022effective}. 
The spontaneous magnetic fields in the superconducting state were explained as a consequence of time-reversal symmetry breaking electron pairing \cite{garaud2014domain,Maiti2015,Vadimov2018,benfenati2020magnetic}. However, the proximity to the mentioned Lifshitz transition and the possibility of another magnetic state at higher doping levels require the investigation of an alternative scenario in which the magnetic fields arise from inhomogeneous spin magnetism with small magnetic moments coexisting with superconductivity.

This scenario can be experimentally investigated by using nuclear magnetic resonance (NMR) as a powerful spectroscopic probe that yields insights into the global and local electronic and magnetic properties of bulk materials. An NMR spectrum represents a histogram of the quasi-static local fields, probed at the positions of the resonantly excited nuclear moments. The relaxation of the excited nuclear moments provides important information regarding the low-energy excitations of the electronic system (fraction of $\mu$eV energy scales). Hence, NMR is a well-established technique to detect the presence of various types of magnetic order, such as spin magnetism. Close to a magnetic transition, spin fluctuations yield a critical slowing down, resulting in an enhancement of the spin-lattice relaxation rate $1/T_1$, with a divergent behavior at the magnetic phase transition. 

Here, we report a detailed $^{75}$As NMR study of \bkfa, with $x$ = 0.77. In contrast to expectations for spin magnetism, we observe a monotonous decrease of $1/T_1T$ across both critical points, $\TcZtwo$ and the superconducting $\Tc{}$. We do not observe an NMR line broadening across these transitions, indicating an extremely weak coupling of spontaneous fields to the nuclear spins. Our NMR data exclude magnetic phases or a proximity to magnetism associated with spin ordering. In contrast, we observe a decrease of $1/T_1T$ below $T^*> \TcZtwo$, indicating a reduction of the low-energy electron density. Our complementary $\mu$SR measurements reveal that the muon Knight shift monotonously decreases below $T^*$ followed by a stronger reduction of the Knight shift below $\Tc{}$, in accordance with the NMR $1/T_1T$ and the NMR Knight-shift data. The observed pseudogap-like behavior is consistent with the formation of bound states of electrons, well above the superconducting phase transition $\Tc{}$. This supports the scenario where the time-reversal symmetry is spontaneously broken by a condensate of pairs of electron pairs above $\Tc{}$ \cite{Grinenko2021state}. 

\begin{figure*}[]
    \centering
    \includegraphics[width=\linewidth]{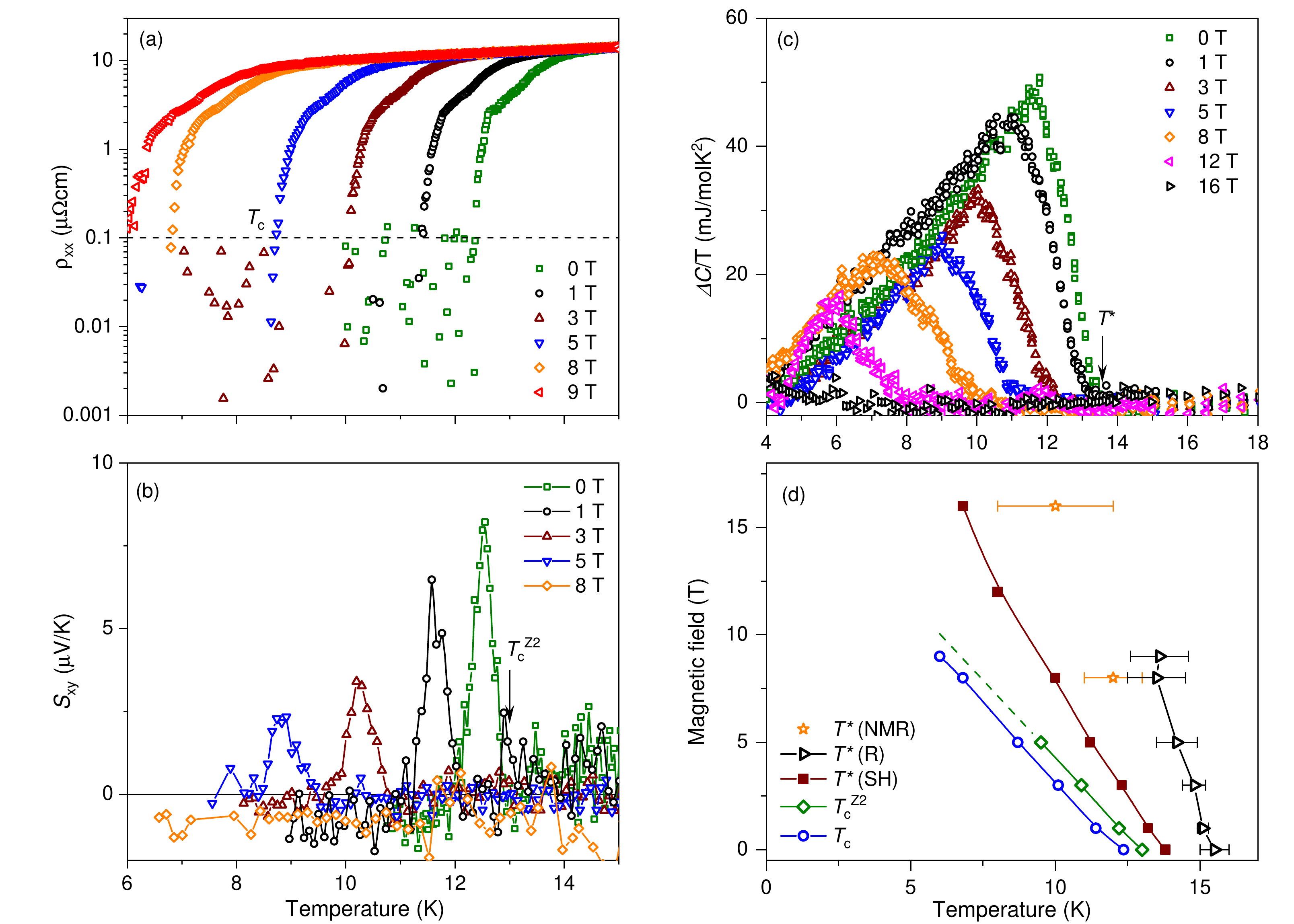}
    \caption{{\bf Physical properties of the \bkfa{} crystal with x = 0.766(1)}. Temperature dependence of (a) the electrical resistivity, $\rho_{\rm xx}$, (b) the spontaneous Nernst effect, $S_{\rm xy}$, and (c) the change of the specific heat across the superconducting transition, $\Delta C/T$, measured at different magnetic fields applied along the $c$ direction. (d) Magnetic field phase diagram of the critical temperatures obtained using various techniques. $\Tc{}$ is the superconducting transition temperature defined by zero resistivity, $\TcZtwo$ is the BTRS transition temperature defined by the onset of the spontaneous Nernst signal, $T^*$(SH) is the onset of the pseudogap extracted from the specific heat, $T^*$(NMR) is obtained from our NMR measurements, presented in Figs.~\ref{fig:NMR_spin-lattice_relaxation_rate_8T} and~\ref{fig:supp_spin-lattice_relaxation_rate_16T}, and $T^*(R)$ is defined as shown in the appendix, Fig.~\ref{fig:rho_pseudogap}. The dashed green line is an extrapolation of $\TcZtwo$ to high fields, where it cannot be resolved in the Nernst measurements anymore.}  
    \label{fig:rxx_Nernst_SH}
\end{figure*}

\section{Characterization by transport and specific heat}
In this work, we measured two \bkfa{} samples: For the NMR measurements, we used a single crystal with $x$ = 0.766(1) and a mass of $m$ = 0.33 mg, and for the $\mu$SR experiments, we used a stack of single crystals with $x$ = 0.75(2) and a total mass of about 12 mg. The magnetic susceptibility of the two samples across the superconducting transition is shown in the appendix in Fig.~\ref{fig:ZFsuscept}. According to our previous studies \cite{Grinenko2021state,shipulin2023calorimetric}, the doping level of these samples falls into the magic doping range, where the electron quadrupling phase and fractional vortices exist \cite{Iguchi2023,zhou2024observation,zheng2024direct}. The temperature dependencies of the electrical resistivity, spontaneous Nernst effect, and specific heat, measured at different applied magnetic fields, are shown in Figs.~\ref{fig:rxx_Nernst_SH}(a)-\ref{fig:rxx_Nernst_SH}(c). The superconducting transition temperature, $\Tc{}$, is defined by zero resistivity using a threshold value of $\rho_{\rm xx} = 10^{-1}\mu\Omega$cm, [Fig.~\ref{fig:rxx_Nernst_SH}(a)], the BTRS transition temperature to the quartic state, $\TcZtwo$ is defined by the onset of a spontaneous Nernst signal, [Fig.~\ref{fig:rxx_Nernst_SH}(b)], and the characteristic crossover temperature for the formation of bound states of electrons, the pseudogap temperature $T^*$(SH), is defined by the onset of the anomaly in the electronic specific heat, [Fig.~\ref{fig:rxx_Nernst_SH}(c)]. Note that, as expected, there is some discrepancy between the value of the crossover temperature $T^*$ measured by different probes, as shown in Fig.~\ref{fig:rxx_Nernst_SH}(d), which is discussed below. 

The overall magnetic-field dependence of the various anomalies (Fig.~\ref{fig:rxx_Nernst_SH}) is qualitatively similar to that observed previously for a sample with a similar doping level \cite{Grinenko2021state}. However, the spontaneous Nernst signal is suppressed somewhat faster with magnetic field in the present sample, and it cannot be resolved at 8 T and beyond. The observed difference is attributed to the slightly different doping levels of the two samples and the high sensitivity of the BTRS phase to K doping~\cite{Grinenko2020superconductivity,shipulin2023calorimetric}. The 
theoretical analysis, presented in the appendix, demonstrates the mechanism of suppression of the quadrupling state when the external magnetic field exceeds a certain value. At the same time, we note that $\TcZtwo$ extrapolated to 8 T is still above $\Tc{}$ as shown in Fig.~\ref{fig:rxx_Nernst_SH}(d). Therefore, using this data alone, we cannot determine whether the quadrupling state is completely suppressed, or if a small quadrupling phase remains at that field with the Nernst signal becoming too small to be resolved in the measurements. In any case, the BTRS transition must be close to $\Tc{}$ at 8 T. The critical temperatures at different fields are summarized in [Fig.~\ref{fig:rxx_Nernst_SH}(d)]. $\Tc{}$ and $\TcZtwo$ closely follow each other, whereas the separation between $T^*$ and those temperatures increases with magnetic field, similar as reported in Ref.~\cite{Grinenko2021state}.

\section{NMR and \texorpdfstring{$\mu$SR}{muSR} data}

\begin{figure}[ht!]
    \centering
    \includegraphics[width=\linewidth]{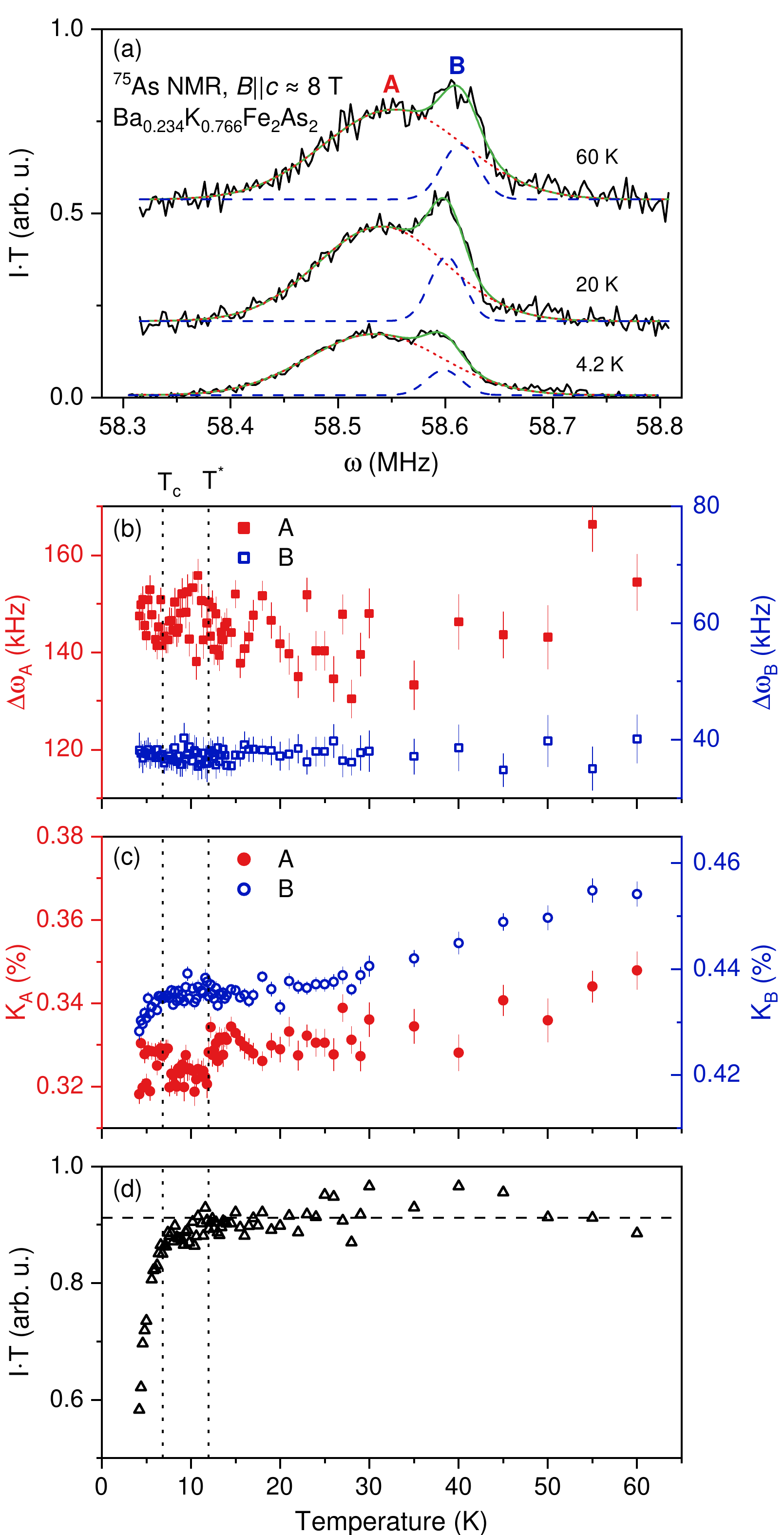}
    \caption{(a) $^{75}$As NMR spectra (black lines), compared to a fit with a double-Gaussian function (green lines), consisting of two peaks A and B (red dotted and blue dashed lines, respectively) at 4.2, 20, and 60 K. The reference frequency is $\omega_0 = 58.350$ MHz. Temperature dependence of (b) the spectral linewidth (rectangles) and (c) the Knight shift (circles) of the peaks A (red full symbols) and B (blue empty symbols). (d) Temperature-dependent intensity of $^{75}$As NMR spectra. The dotted vertical lines in (b)-(d) indicate $\Tc{}$ and $T^*$. The dashed horizontal line in (d) is a guide to the eye, indicating the expected behavior of a simple metal.}
    \label{fig:NMR_spectrum}
\end{figure}

In \Figref{fig:NMR_spectrum}(a), we show the temperature-normalized $^{75}$As NMR spectra of \bkfa{} with $x$ = 0.766(1) at 60, 20, and 4.2 K in 8 T applied along the $c$ axis (for sample orientation and confirmation of the bulk potassium stoichiometry see Figs. \ref{fig:supp_NMR_orientation} and \ref{fig:supp_quadrupole_frequency} in the appendix, respectively). 
For all temperatures, the spectra can be well described by two Gaussian contributions, a broader peak (A)  at lower frequencies and a sharper peak (B) at higher frequencies [red and green dashed lines in \Figref{fig:NMR_spectrum}(a)], respectively.

We attribute these to the presence of nanoclusters of potassium-rich and potassium-poor regions in the sample.
We performed scanning tunneling microscopy (STM) measurements and found evidence for such a separation, at least near the surface (see Fig.~\ref{fig:STM} in the appendix). These nanoclusters are a local phenomenon and do not affect the global electronic properties probed by NMR, as will be discussed in the following.

Next, we discuss the spectral properties. From 60 to 20 K, the peak intensity $I$ (proportional to the nuclear magnetization of the contributing sample volume) of both peaks A and B remains unchanged. However, at 4.2 K, below $\Tc{}$, the intensity of both peaks decreases. This indicates a typical metallic behavior ($I \propto 1/T$) in the high-temperature range. Below $\Tc{}$, the contributing volume fraction of nuclear spins decreases due to radio-frequency (RF) shielding of the NMR pulse in the superconducting state, leading to a decrease in signal intensity. Figure \ref{fig:NMR_spectrum}(d) shows the temperature-normalized intensity of all recorded $^{75}$As NMR spectra, which is obtained by integration over frequency of the entire spectrum at a given temperature. At high temperatures, the intensity is constant within error bars. Exactly at $\Tc{}$, the RF shielding causes a sharp drop in intensity. 
Upon closer inspection, the intensity is already slightly reduced in the temperature regime $T^* > \Tc{}$. We can interpret this as microscopic RF shielding due to the appearance of non-condensed Cooper pairs, which also results in a gradual reduction of the electrical resistivity below $T^*$ [Fig.~\ref{fig:rho_pseudogap}]. The indication of the presence of non-condensed pairs has important phenomenological consequences. It serves as a prerequisite for electron quadrupling condensation through fluctuation-based mechanisms \cite{bojesen2014phase,babaev2004superconductor}.

In order to obtain quantitative information on the spectral linewidth $\Delta \omega$ and Knight shift $K$, a separate analysis of the contributing peaks A and B is necessary.

In \Figref{fig:NMR_spectrum}(a), we compare the $^{75}$As NMR spectra with a fit of two overlapping Gaussian peaks, which describes the spectra well for all temperatures. From the fits, we extract $\Delta \omega$ and the line position (Knight shift) of the single peaks and plot these in Figs.  \ref{fig:NMR_spectrum}(b) and \ref{fig:NMR_spectrum}(c), respectively. The results for peak A exhibit greater scattering compared to peak B. However, the spectral properties of both contributions show the same qualitative behavior across the whole temperature range.  

The spectral linewidth is a direct measure of the local-field distribution. Given the Gaussian shape of both contributions to the NMR spectrum, we can calculate the linewidth directly from the second spectral moment $M_2$ as $\Delta \omega \propto \sqrt{M_2}$. Consequently, any change of the underlying distribution of hyperfine fields will be reflected in a change of $\Delta \omega$. 

From the data in \Figref{fig:NMR_spectrum}(b), we do not find any temperature dependence of $\Delta \omega$ when crossing $\TcZtwo$, indicating that the time-reversal symmetry breaking occurs in the absence of magnetic order, and, hence, is of nonmagnetic origin.

\subsection{NMR Knight shift}

In order to probe the local spin susceptibility in the superconducting state, measurements of the NMR Knight shift are a well-established experimental technique. The Knight shift can be written as $K = K_s + K_o$, where $K_s$ denotes the local spin susceptibility and $K_o$ stems from orbital moments. Typically, $K_o$ is temperature-independent and contributes as an offset to the total Knight shift. The temperature dependence of $K$ is primarily determined by the spin part $K_s$. The latter is directly proportional to the spin part of the electronic susceptibility $\chi_s$, and can be written as 
\begin{equation}
    K_s \propto \chi_s = -4\mu_B^2 \int_0^{\infty}N(E) \frac{\partial f(E)}{\partial E} dE, 
\end{equation}
with the Fermi-distribution function $f(E)$ and the electronic density of states $N(E)$.

The temperature-dependent Knight shift can potentially provide information about the superconducting gap structure. As a well-known example, for a fully gapped superconducting state, an exponential decrease of the local spin susceptibility is expected below $\Tc{}$, as described by the Yosida function \cite{yosida_paramagnetic_1958}. For nodal gap structures, a power-law behavior is expected. Finally, in the case of a triplet-pairing state, where cooper pairs have a total spin of 1, the local spin susceptibility would remain constant below $\Tc{}$ for certain directions.

As shown in \Figref{fig:NMR_spectrum}(c), the temperature-dependent Knight shift drops below $\Tc{}$, consistent with expectations for a singlet state. A low-temperature power-law behavior may be expected due to the rather complicated momentum dependence of the superconducting gap over the Fermi surface observed for the neighboring doping levels~\cite{Xu2013}. 
However, our measurements were only performed down to 4.2 K, which is not low enough to reliably determine the gap structure. 

At high temperatures, the Knight shift changes monotonously, contrary to expectations for a simple metal, which would follow the Korringa relation ($K_s^2 T_1 T = \text{const.}$) \cite{korringa_nuclear_1950}. Grafe \textit{et al.}  reported a similar behavior for another iron-based superconductor, LaFeAsO$_{0.9}$F$_{0.1}$, and associate this with a gapped behavior, comparable to that observed in cuprates \cite{grafe_as_2008}. In the range between $\Tc{}$ and $T^*$, the Knight shift is constant within the experimental error and yields no discernible features, as is seen best for the results from peak B. To obtain additional information regarding the static spin susceptibility below $T^*$, we performed high-resolution $\mu$SR Knight-shift measurements.  

\subsection{\texorpdfstring{$\mu$SR}{muSR} Knight shift}
\begin{figure}[ht!]
    \centering
    \includegraphics[width=\linewidth]{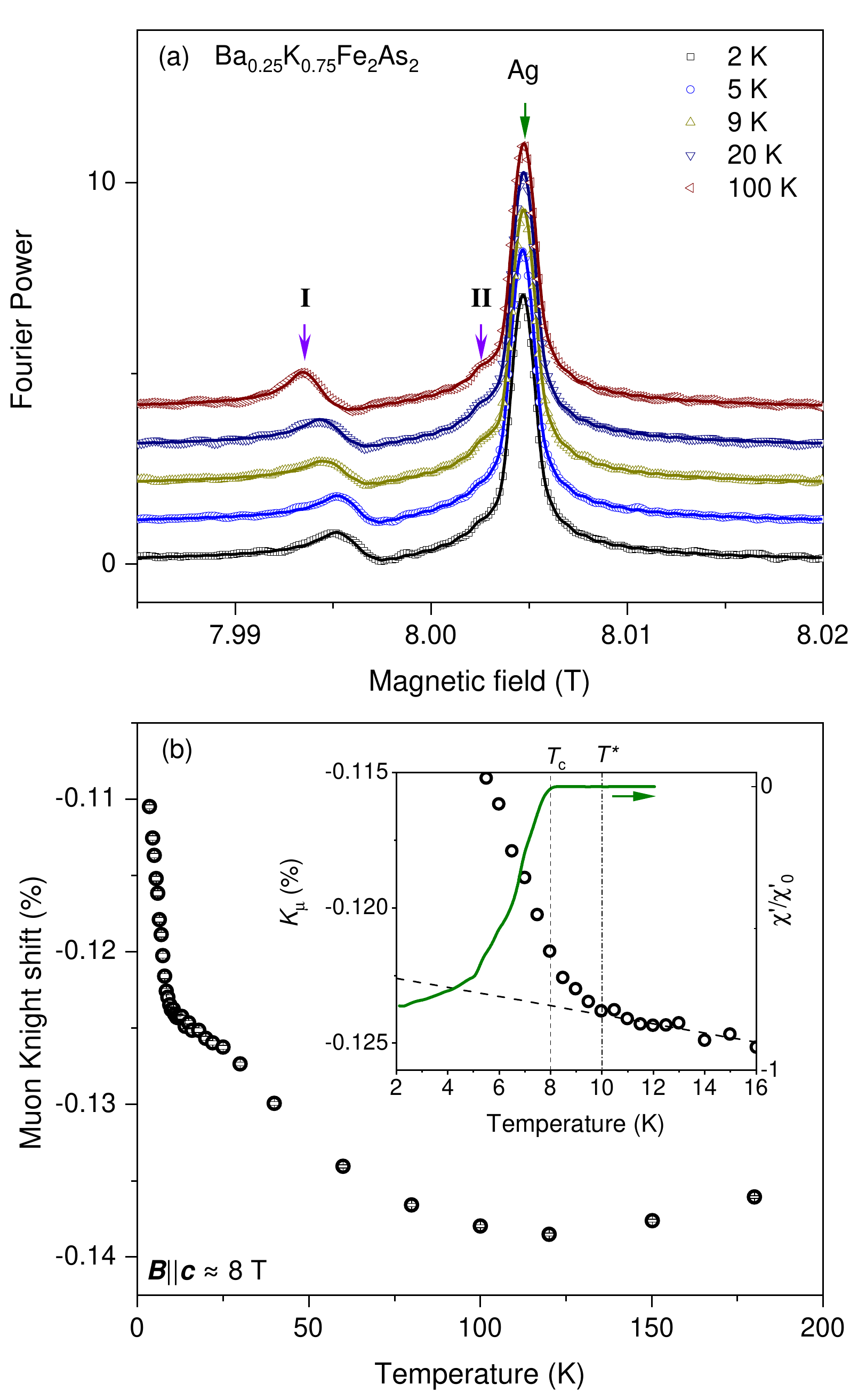}
    \caption{(a) Fourier transform of the TF $\mu$SR time spectra measured at about 8 T and selected temperatures for \bkfa{} with $x$ = 0.75(2). The arrows indicate different components of the signal: \RNum{1} and \RNum{2} are muons stopping at two different sites of the sample, and Ag denotes a signal from muons stopping in the silver sample holder. (b) Temperature dependence of the $\mu$SR Knight shift measured at about 8 T, applied along the $c$ axis. The inset shows data close to the superconducting transition. The negative Knight shift (left) becomes less negative already somewhat above $\Tc{}$ indicating the formation of Cooper pairs above the superconducting transition. We determined $\Tc{}$ via AC susceptibility at 8 T (using an AC field of 10 $\mu$T at a frequency of 417 Hz). The data are normalized by $\chi_0$, measured at 2 K and in zero field. The muon Knight shift deviates from the normal state behavior at the pseudogap temperature  $T^*$($\mu$SR)$~>~\Tc{}$.}
    \label{fig:muSR}
\end{figure}

We performed transversal-field (TF) $\mu$SR measurements on the stack of \bkfa{} single crystals with $x$ = 0.75(2) at 8 T. Several examples of fast Fourier transform high-field $\mu$SR time spectra, measured at different temperatures, are shown in Fig.~\ref{fig:muSR}(a). The spectra consist of three peaks. Similar to previous observations for FeSe \cite{Grinenko2018}, there are two muon-stopping sites, \RNum{1} and \RNum{2}, with a more intense \RNum{1} peak. In addition, the peak \RNum{2} overlaps with the Ag holder background, which makes the analysis of its temperature dependence less reliable. Therefore, we focused on the analysis of peak \RNum{1} relative to the Ag peak, which within error bar of the experiment has a temperature-independent Knight shift \cite{Grinenko2018}.

To obtain the Knight shift, we described the spectra using three $A_i\text{cos}(\omega_it+\phi)\text{exp}(-\lambda_it)$ components, where $A_i$ corresponds to the muon fractions stopped in the Ag holder and at the \RNum{1} and \RNum{2} sites. The fractions were defined at the base temperature and fixed for all other temperatures. Further details of the analysis can be found in Ref.~\cite{Grinenko2018}. The resulting temperature dependence of the Knight shift, defined as $K_\mu = (B_{\rm \RNeq{1}}-B_{\rm Ag})/B_{\rm Ag}$, is plotted in Fig.~\ref{fig:muSR}(b), where $B_{\rm \RNeq{1}}$ is the temperature-dependent position of peak \RNum{1} and $B_{\rm Ag}$ is the position of the Ag peak. The Knight shift was not corrected for the demagnetizing effect, which is assumed to be significant for the present sample geometry with an area of about $4 \times 4$ mm$^2$ and effective thickness (accounting for the gaps between the crystals) of 300 $\mu$m. In the following, we assume that the demagnetizing effects do not affect the temperature dependence above $\Tc{}$ qualitatively.  

In the normal state, the Knight shift is negative, similar to the one measured for FeSe \cite{Grinenko2018}. The overall temperature dependence is consistent with the normal-state susceptibility, showing a broad maximum at around 100 K for this doping level \cite{Grinenko2020superconductivity}. At lower temperatures, the negative Knight shift reduces linearly towards $\Tc{}$, without any sign of a Curie-Weiss behavior, as would be expected in the case of local magnetic moments [see inset in Fig.~\ref{fig:muSR}(b)]. The Knight shift deviates from the linear, normal-state behavior at $T^*$, which is about 2 K above $\Tc{}$, measured by AC susceptibility at 8 T. The Knight shift shows a second kink at $\Tc{}$. The presence of two features is qualitatively consistent with the temperature dependence of the NMR intensity [Fig.~\ref{fig:NMR_spectrum}(d)]. The smaller splitting between $\Tc{}$ and $T^*$ observed for the $\mu$SR sample is explained by the smaller potassium doping level~\cite{shipulin2023calorimetric}. To conclude, the NMR and $\mu$SR Knight-shift data consistently indicate the absence of local magnetic moments and magnetic phase transitions in \bkfa{} at around the magic doping level. The overall data are consistent with the existence of a pseudogap above $\Tc{}$, with $T^*$ being highly sensitive to the K doping level. We did not find any resolvable feature in the Knight shift in the vicinity of $\TcZtwo$, which is consistent with the scenario of the quartic state, where the BTRS transition originates from the ordering of relative phases of the pairing fields in different bands.  

\section{Spin-lattice relaxation rate}

Since the Knight shift probes the static susceptibility only at $q=0$, we also need to investigate the spin-lattice relaxation rate, which probes the dynamic susceptibility over the whole $q$ space and is sensitive to low-energy spin fluctuations, which are expected to yield a significant increase close to a magnetic instability.

\begin{figure}[ht!]
    \centering
    \includegraphics[width=\linewidth]{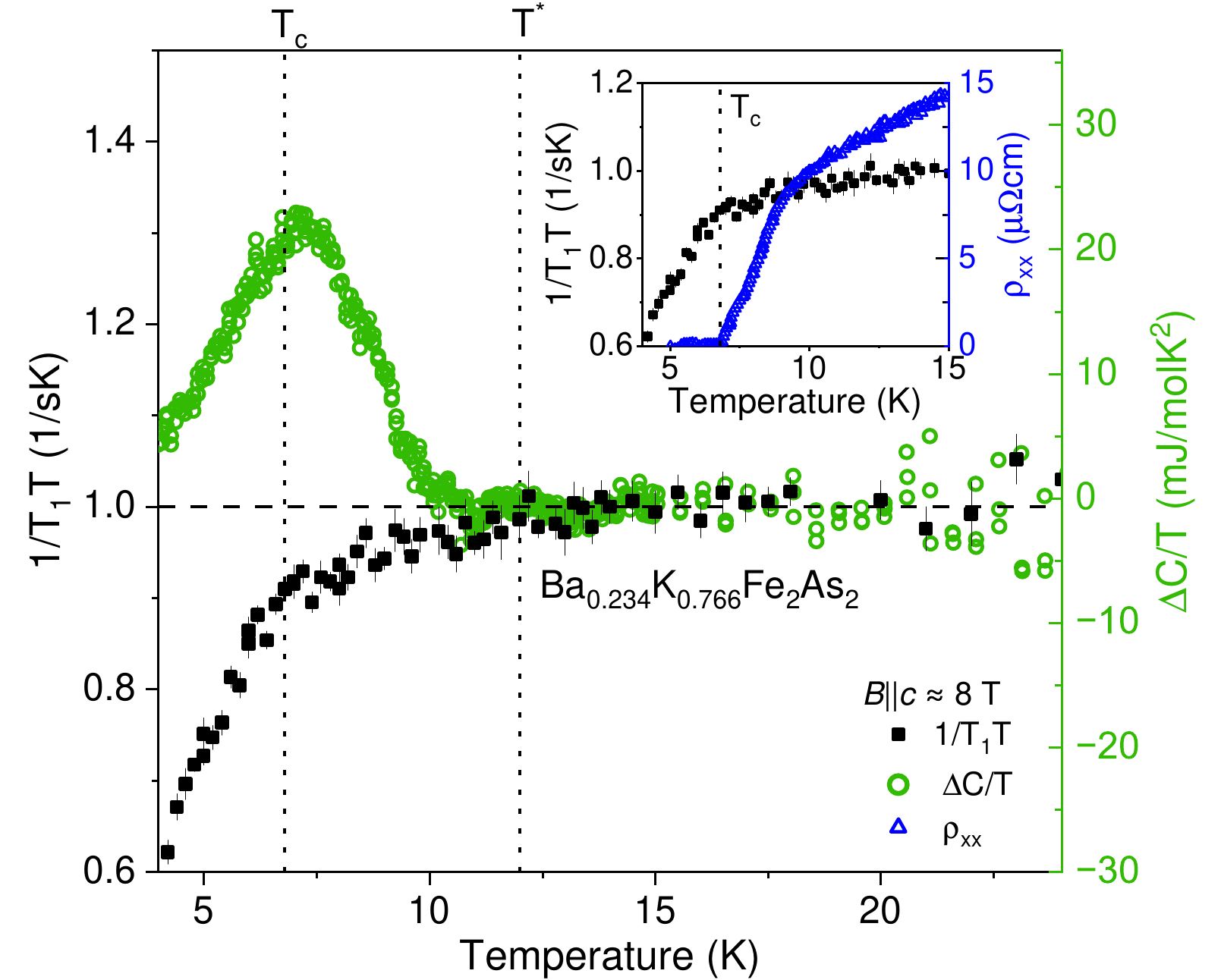}
    \caption{Temperature dependence of the $^{75}$As nuclear spin-lattice relaxation rate $1/T_1T$ (black rectangles), the specific-heat difference divided by $T$ (green circles), and the magnetoresistance in the inset (blue triangles) at 8 T. The dotted, vertical lines mark the critical temperatures $\Tc{}$ and $T^*$(NMR).
    The horizontal dashed line is a guide to the eye. The monotonous reduction of $1/T_1T$ below $T^*$(NMR) also indicates that the pronounced anomaly in the specific heat is not related to a magnetic transition. The resistivity shows a gradual deviation from the normal-state behavior slightly above $T^*$(NMR), as can be seen in Fig.~\ref{fig:rho_pseudogap} of the appendix.}
    \label{fig:NMR_spin-lattice_relaxation_rate_8T}
\end{figure}

The nuclear spin-lattice relaxation rate $1/T_1T$ is proportional to the $\vec{q}$-dependent dynamic susceptibility,
\begin{equation}
    \frac{1}{T_1 T} \propto \sum_{\vec{q}} |A_{\perp}(\vec{q})|^2 \frac{\chi_{\perp}''(\vec{q}, \omega_L)}{\omega_L},
\end{equation}
with the transverse components of the hyperfine coupling $A_{\perp}(\vec{q})$ and the transverse imaginary part of the dynamic susceptibility at the nuclear Larmor frequency $\chi_{\perp}''(\vec{q}, \omega_L)$.
We show the temperature-dependent $^{75}$As nuclear spin-lattice relaxation rate of \bkfa{} with $x=0.776(1)$ at 8 T and compare it to the temperature dependence of the specific heat and electrical resistivity at the same field in \Figref{fig:NMR_spin-lattice_relaxation_rate_8T}.

At temperatures between 12 and 25 K, $1/T_1T$ is constant within error bars, which is expected for a normal-state metallic behavior. We note that at temperatures above about 25 K, $1/T_1T$ increases monotonously (see \Figref{fig:supp_high_T_pseudogap_8T} in the appendix). 

$1/T_1T$ begins to decrease below $T^* \approx 12$ K, and this trend continues down to $\Tc{}$. In this regime, there are no signs of a spin-fluctuation-driven enhancement. The specific heat starts to increase between $\Tc{}$ and $T^*$, with an onset temperature far above $\Tc{}$. Grinenko \textit{et al.} \cite{Grinenko2021state} observed the separation of the onset temperature of the specific-heat anomaly, here denoted as $T^*$(SH), and $\Tc{}$, measured by zero resistance. This behavior is consistent with the specific heat measured in our sample.

We identify the slope change of $1/T_1T$ with the formation of a bound state of electrons (non-condensed Cooper pairs, that form prior to the quadrupling transition and coexist with the quadrupling order parameter below $\TcZtwo$) well above the superconducting transition temperature $\Tc{}$.
This conclusion was reached in Ref.~\cite{Grinenko2021state}, based on complementary electric and thermoelectric transport data. However, the direct spectroscopic properties of these states and direct evidence for a pseudogap have been missing so far. Importantly, since we address a crossover, there are no singularities at $T^*$. Instead, the characteristic temperature $T^*$ is an approximate quantity. Hence, the various experimental probes exhibit different sensitivities and result in slightly different $T^*$ values. We find that at 8 T, $1/T_1T$ deviates from the normal-state behavior about 2 K above the onset of the anomaly in the specific heat, and this separation becomes even larger in higher fields, compare Figs.~\ref{fig:rxx_Nernst_SH}(d) and~\ref{fig:supp_spin-lattice_relaxation_rate_16T}. This discrepancy is expected, in particular, due to the different time scales of the experimental probes. 

At $\Tc{} = 6.8$ K, where the resistance becomes zero, we observe a sharp drop in $1/T_1T$ without the appearance of a Hebel-Slichter peak (Fig.~\ref{fig:NMR_spin-lattice_relaxation_rate_8T}), which is a commonly observed behavior for iron-based superconductors \cite{Zhang2010,Zhang2018,Carretta2020}. 

We note that our $1/T_1T$ data of the $x=0.776(1)$ sample fit very well into the systematic behavior reported for \bkfa{} with different K-doping levels \cite{hirano_potential_2012} (see \Figref{fig:supp_spin-lattice_relaxation_rate_Hirano}).
 
Finally, we conclude that the overall behavior of $1/T_1T$ does not reveal any noticeable enhancement of low-energy spin fluctuations close to $\Tc{}$. If such an enhancement were present, it would raise the question if the breaking of time-reversal symmetry would be related to magnetic ordering. To conclude, the temperature dependence of $1/T_1T$, combined with the NMR and $\mu$SR Knight shift data, provides convincing evidence for the absence of spin-related magnetism in \bkfa{} at the magic doping level.

\section*{Discussion}
We characterized \bkfa{} by means of NMR and $\mu$SR. This allowed us to examine the possible presence and contribution of magnetic effects at the magic doping level. Our results are consistent with previous experiments, which were interpreted as evidence for the formation of a new state of matter above the superconducting phase transition. The combined results show that above $\Tc{}$, there is a phase fluctuation-driven electron quadrupling condensate \cite{Grinenko2021state, shipulin2023calorimetric}.

The previously observed spontaneous Nernst effect~\cite{Grinenko2021state} is gradually
suppressed by an application of an external field. We showed here that the effect has a non-magnetic origin. We demonstrate in the Methods sections that, instead, it can be explained within the electron quadrupling condensate scenario, implying that the BTRS state is associated with interband phase locking of the multicomponent superconducting order parameter. In this scenario, a pseudogap appears at $T^*$, defined as a crossover temperature for the onset of pairing correlations, preceding the onset of four-electron correlations and eventual electron quadrupling condensation. The critical temperature, $\TcZtwo$ ($T^* > \TcZtwo > \Tc{}$), is associated with the time-reversal-symmetry-breaking ordering of the phase difference of fluctuating pairing fields belonging to different bands $\Delta_a$ and $\Delta_b$. In this BTRS state, the order parameter is fourth-order in fermionic fields $\expval{ \Delta_a\Delta_b^*} \ne 0$, while there are no broken symmetries at the level of electron pairing $\expval{ \Delta_a } =0$. Hence, the state represents fermion quadrupling. The superconducting transition at lower temperatures is associated with the onset of long-range order in the individual fields $\expval{ \Delta_a }$.  

\subsection{Assessment of  magnetic-ordering scenarios}
The obtained experimental data provide evidence against other scenarios in which the BTRS state is associated with a magnetic transition or the existence of local magnetization or magnetic/superconducting phase separation. Ferromagnetic order is expected to be enhanced by the applied magnetic field. This is in conflict with the opposite field dependence of the BTRS phase transition. This magnetism-based scenario is also inconsistent with the suppression of the spin susceptibility below $T^*$, probed by the NMR and $\mu$SR Knight shifts, see Figs.~\ref{fig:NMR_spectrum} and~\ref{fig:muSR}. 
In principle, an antiferromagnetic state can result in a spontaneous Nernst effect, as discussed in Ref.~\cite{Grinenko2021state}, which can be suppressed by a magnetic field. However, the suppression of the antiferromagnetic order would enhance the magnetization, which is in conflict with the observed reduction of the Knight shift at low temperatures. 

Also, our data exclude proximity to magnetism since we did not observe any noticeable enhancement of the NMR spin-lattice relaxation rate at low temperatures, which could be related to a slowing down of the spin fluctuations close to a magnetic transition. This provides evidence that spontaneous magnetic fields detected by zero-field $\mu$SR originate from persistent real space currents, which is one of the characteristics of BTRS electron quadrupling state. 

\subsection{Electronic spectral characteristics}

Our NMR experiments demonstrate nontrivial spectral characteristics above the superconducting phase transition: The existence of a pseudogap. We find that the pseudogap formation correlates with the change in transport properties. In \bkfa{} at magic doping, the resistivity decreases below $T^*$ as shown in Fig.~\ref{fig:rho_pseudogap} and also in Ref.~\cite{Grinenko2021state}, despite the decrease in the density of state as indicated by the NMR measurements. We interpret the reduction of the resistivity and the consistent formation of the NMR pseudogap by originating in the formation of incoherent Cooper pairs. The formation of non-condensed Cooper pairs at temperatures preceding the phase transitions is a necessary ingredient of the formation of an electron quadrupling condensate via the phase fluctuation mechanism discussed in \cite{Grinenko2021state,bojesen2014phase,Bojesen2013time,Maccari2022effects}.

Hence, our overall conclusion is that the unconventional properties of \bkfa{} at magic doping originate from electron pairing and quadrupling in the absence of magnetism.

It is worth mentioning that a pseudogap ~\cite{Shi2018, Kang2022} and BTRS superconductivity ~\cite{Matsuura2023} were reported recently as well for the Fe(Se,S,Te) system. This calls for investigations whether a quartic state may exist at certain doping levels also in this system. A promising direction is to search for quartic states in low-dimensional systems and monolayers of BTRS superconductors, in which the fluctuations are enhanced by reduced dimensionality. In particular, a pseudogap temperature exceeding twice the resistive $\Tc{}$ was reported in single-layer FeSe/SrTiO$_3$~\cite{Faeth2021} that calls for further investigation. Another realization of the quadrupling phase is expected on the $1 \times 1$ potassium-terminated surface of KFe$_2$As$_2$ \cite{zheng2024direct}. STM data indicate that this surface is a 2D superconductor with broken time-reversal symmetry. Further studies are needed to investigate a possible pseudogap behavior and to search for a quadrupling phase in this novel 2D superconductor.

\section*{Methods}

\subsection*{Samples}
The K doping level $x$ of the single crystal used for the NMR measurements was determined using the known relation between the $c$-axis lattice parameter and the K doping, determined in previous studies: $x = -14.574 + 1.1234c({\rm \AA})$, $R$ = 0.99852 \cite{Kihou2016}. To confirm that the sample is single phase, we performed 2$\Theta$ x-ray measurements from both sides of the crystal, see Fig.~\ref{fig:supp_XRD}. Here, we obtained the same $c$-axis parameter within 0.01$\%$, indicating a homogeneous distribution of the K doping in the NMR sample. For the $\mu$SR measurements, we selected about 30 crystals based on the $\Tc{}$ obtained in low-field magnetization measurements using Quantum Design MPMS and PPMS devices. The resulting real part of the AC susceptibility in zero static (DC) magnetic field is shown in Fig.~\ref{fig:ZFsuscept}. The $\mu$SR sample has a slightly higher $\Tc{}$ and a broader transition width compared to the NMR sample. Using the $\Tc{}$ value and observed transition width, we estimated the doping level of the $\mu$SR sample, $x$ = $0.75(2)$ K, using the previously obtained phase diagram from Ref.~\cite{Grinenko2020superconductivity}.

\subsection*{Experimental}
The electrical resistivity, Nernst- and Seebeck-effect measurements of the NMR sample were performed using a homemade sample holder for transport properties, inserted in a Quantum Design PPMS equiped with a 9 T magnet. The Seebeck coefficient ($S_{\rm xx}$) is the ratio of the longitudinal electric field to the longitudinal thermal gradient applied to generate it. The Nernst coefficient ($S_{\rm xy}$) is related to the transverse electric field produced by a longitudinal thermal gradient. The procedure to obtain a spontaneous Nernst signal is described in Refs.~\cite{Grinenko2021state,shipulin2023calorimetric}. The low-temperature specific heat of the NMR sample at different magnetic fields was measured in a Quantum Design PPMS, using the thermal-relaxation method. The transversal-field (TF) $\mu$SR measurements were performed at the HAL-9500 spectrometer at the Swiss Muon Source (S$\mu$S), PSI, Villigen, in an 8 T magnetic field applied along the $c$ axis.

The NMR measurements were performed in a $^4$He flow cryostat, mounted in an 8 T superconducting magnet with a spatial field homogeneity of about 1 ppm over the sample volume. The $^{75}$As NMR data were recorded using a commercial phase-coherent spectrometer. We used standard Hahn spin-echo pulse sequences to measure the spectra, with carefully adjusted pulse parameters in order to avoid RF heating effects. We measured the nuclear spin-lattice relaxation time $T_1$ by using a saturation-recovery method, and described the data using the  stretched-exponential function $m(t) = m(\infty) [1 - A (0.1 e^{-(t/T_1)^{\beta}} + 0.9 e^{-(6t/T_1)^{\beta}})]$ for the central transition ($I_z = -1/2 \leftrightarrow +1/2$), where $\beta$ is a stretching exponent, $m(\infty)$ is the nuclear magnetization at equilibrium ($t \rightarrow \infty$), and $A$ is the inversion factor. The superconducting phase diagram of the NMR sample was checked by monitoring the complex RF reflection coefficient S11 at the NMR circuit, using a vector network analyzer (\Figref{fig:supp_resonant_freq}).

The STM/S experiments were performed using a commercial Unisoku USM1300 low-temperature STM. Pt/Ir tips were used, and conditioned by field emission with a gold target. 
To obtain a clean surface for the STM measurements, the sample \bkfa{} with $x$ = 0.771(2) was cleaved in-situ at temperature of $78\,\mathrm{K}$ in an ultrahigh vacuum (base pressure $\approx 2\times 10^{-10}$ mbar), then transferred immediately to the STM stage (maintained at $4.2\,\mathrm{K}$) for STM/S measurements. The magnetic susceptibility data for this sample is shown in Fig.~\ref{fig:ZFsuscept}, and other physical properties are presented in Ref.~\cite{Grinenko2021state}.

\subsection*{Analysis of field dependence of BTRS transition}

Two additional observations are consistent with the interpretation that the time-reversal symmetry breakdown is due to the onset of electron quadrupling order, but not with magnetism. The first aspect is that the critical temperature $\TcZtwo$ is suppressed by external field. This effect occurs because the quadrupling transition is associated with the onset of an interband relative phase in a system of preformed Cooper pairs. The external magnetic field breaks Cooper pairs and, hence, suppresses the phase-ordering temperatures both for the quadrupling (interband relative phase ordering at $\TcZtwo$) and the superconducting transition  (interband phase sum ordering at $\Tc{}$). The breakdown of time-reversal symmetry involves two inequivalent lockings of the interband relative phase, associated with the same energy \cite{Bojesen2013time,bojesen2014phase,Grinenko2021state}. Both phase lockings are related to each other by time-reversal operation. This spontaneous breakdown of the time-reversal $\Ztwo$ symmetry involves relative phase degrees of freedom and, thus, is not a conventional breaking of time-reversal symmetry as in magnetism. In particular, the external magnetic field does not break explicitly this $\Ztwo$ symmetry. In other words, the relative phases do not have a Zeeman-type coupling to external field, such as magnetic spins would have. 

The second observation is that a sufficiently strong magnetic field eliminates the BTRS state, or at least makes the spontaneous Nernst effect undetectable in our measurement. To understand whether the observed suppression of the spontaneous Nernst signal in an external magnetic field is consistent with the BTRS superconductivity and electron quadrupling, we performed finite-element simulations within Ginzburg-Landau theory.
Our simulations, based on a minimal microscopic $s+is$ model demonstrate that, for the considered range of parameter, the external magnetic field suppresses $\TcZtwo$ associated with time-reversal-symmetry-breaking interband phase locking (see details in appendix). 

\appendix

\section{Sample characterization}
\subsection{X-ray measurements}
\begin{figure}[ht!]
	\centering
	\includegraphics[width=\linewidth]{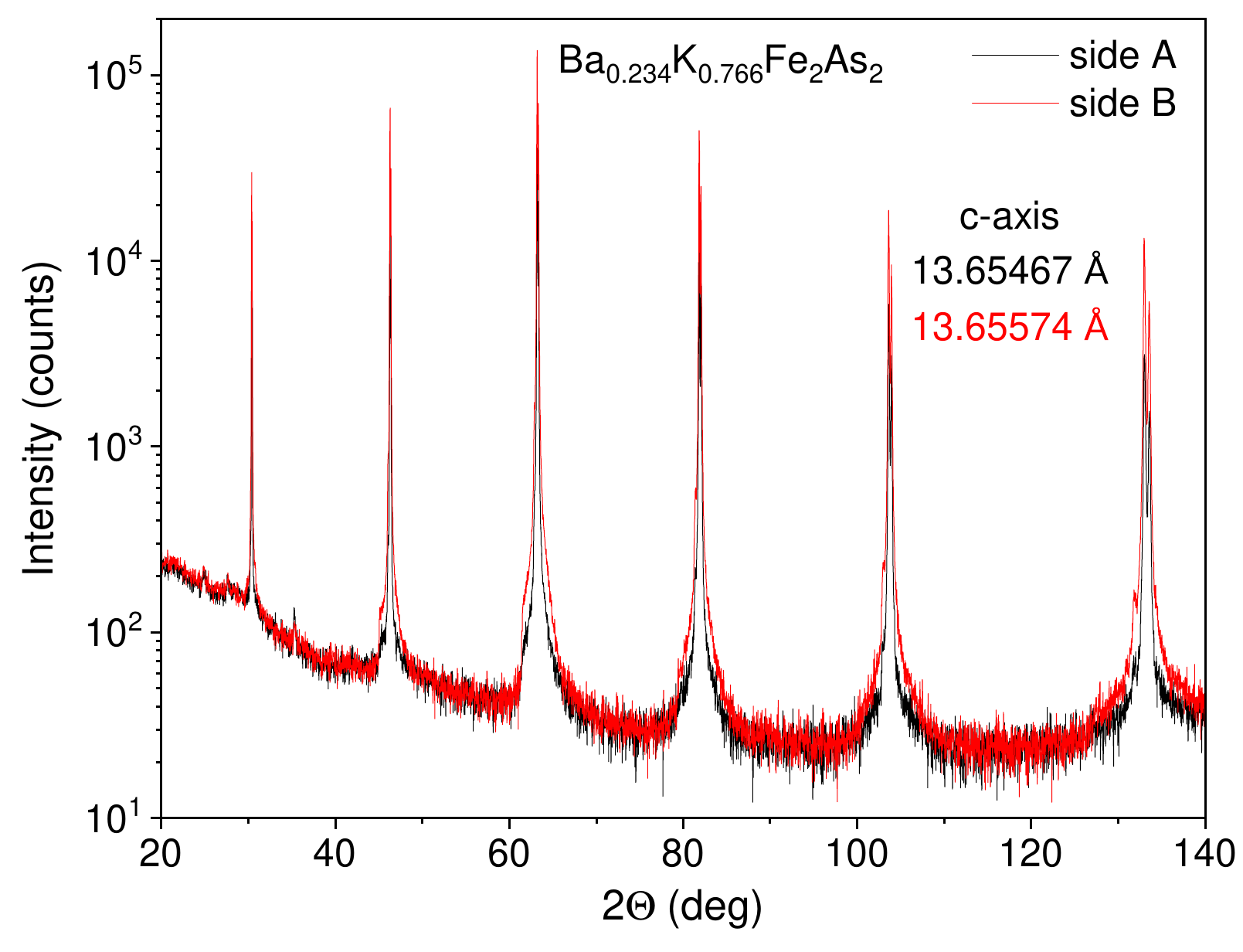}
	\caption{2$\Theta$ X-ray scans measured from both sides of the sample used for the NMR measurements. The obtained $c$-axis parameter corresponds to x = 0.766(1) \cite{Kihou2016}, indicating that the doping level is the same within 0.1\% for both sides of the sample.}
	\label{fig:supp_XRD}
\end{figure}
We performed X-ray measurements on both sides of the NMR sample (Fig.~\ref{fig:supp_XRD}). The $c$-axis parameter corresponds to a doping level of $x = 0.766(1)$ for both sides defined using the relation between the K doping level and the $c$ axis obtained in the previous works \cite{Kihou2016,shipulin2023calorimetric}. This confirms a high homogeneity and uniform distribution of Ba/K within the sample.
\newpage
\subsection{Susceptibility}
\begin{figure}[ht!]
	\centering
	\includegraphics[width=\linewidth]{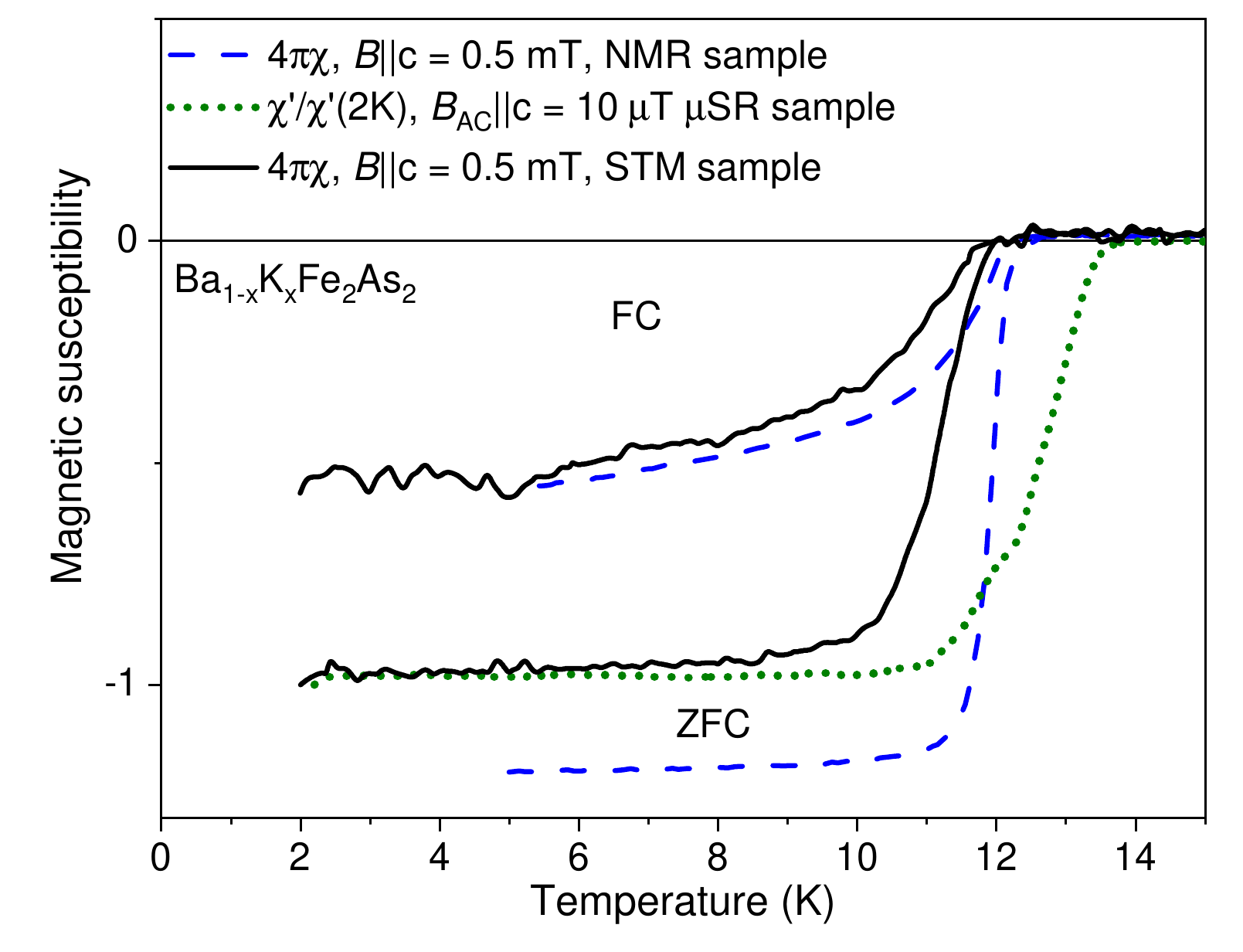}
	\caption{Temperature-dependent magnetic susceptibility of the single crystals used for the NMR and STM measurements as well as of the stack of the single crystals used for $\mu$SR measurements (dashed blue, solid black, and dotted green line, respectively).}
	\label{fig:ZFsuscept}
\end{figure}
Fig. \ref{fig:ZFsuscept} compares the low-field susceptibility of the NMR, $\mu$SR, and STM sample. According to the previously determined phase diagram \cite{Grinenko2021state}, $\Tc{}$ of all the samples falls into the magic doping range. $\Tc{}$ of the $\mu$SR sample (stack of single crystals) is slightly higher than $\Tc{}$ of the NMR and STM samples (individual single crystals) and corresponds to a doping level of $x=0.75(2)$. The broader transition of the $\mu$SR sample can be attributed to slightly different doping levels of the individual crystals in the stack.

\newpage
\section{NMR}

\subsection{Sample orientation}

\begin{figure}[ht!]
	\centering
	\includegraphics[width=\linewidth]{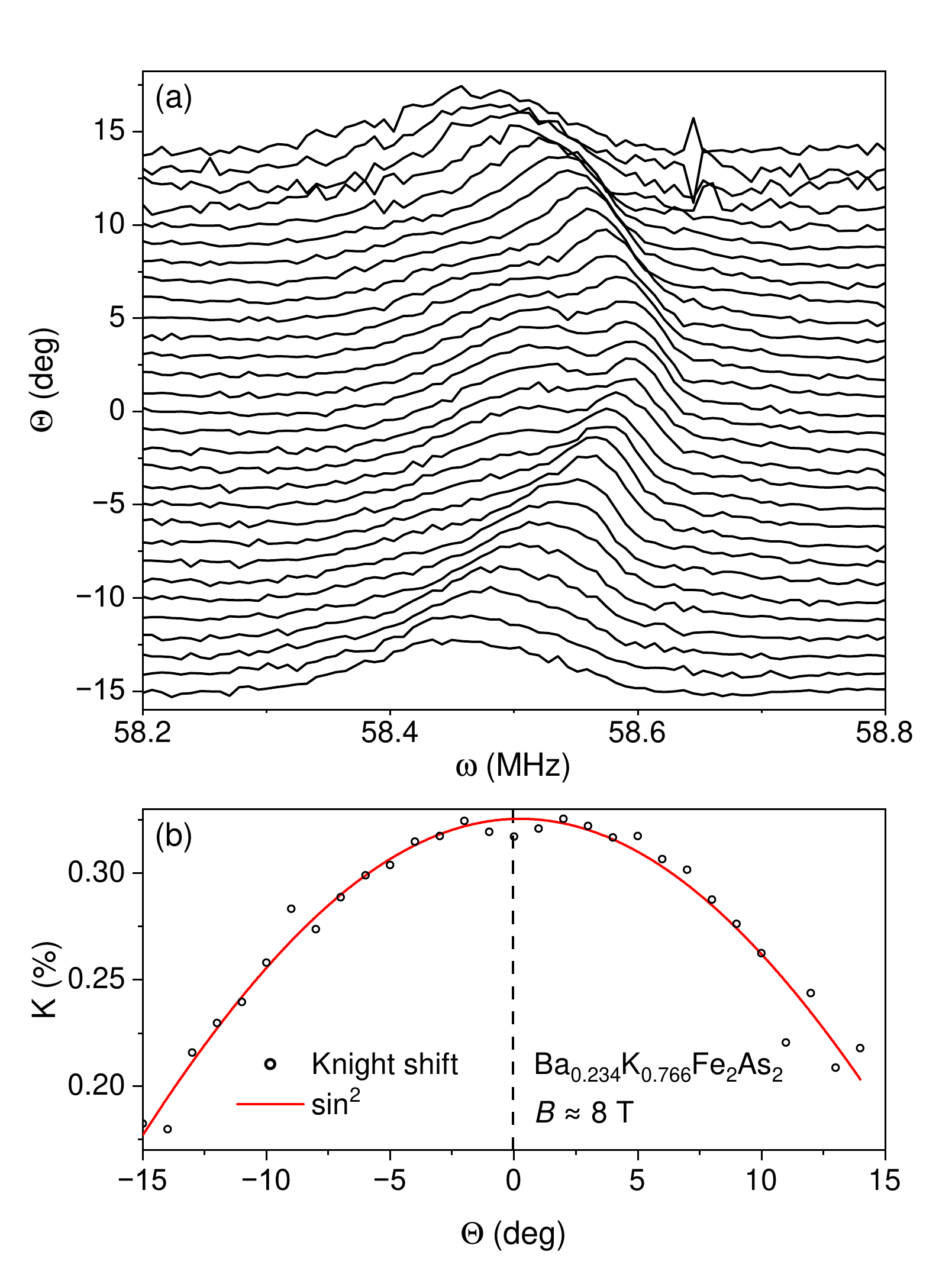}
	\caption{(a) Angular dependence of the $^{75}$As NMR spectra at 8 T and 10 K. (b) Angular dependence of the Knight shift (black circles), determined from the spectra in (a). The Knight shift is compared to a fit with a squared-sine function (red line). The maximum of this fit corresponds to a field orientation parallel to the crystallographic $c$ axis.}
	\label{fig:supp_NMR_orientation}
\end{figure}

For the NMR measurements, we utilized the angular dependence of the $^{75}$As Knight shift to orient the \bkfa{} sample in magnetic field. At 10 K and 8 T, the angular-dependent shift yields a squared-sine-type behavior, where the maximum of the shift corresponds to a field orientation parallel to the crystallographic $c$ axis. \Figref{fig:supp_NMR_orientation} shows the angular-dependent NMR spectra over a range of $\pm 15$° [\Figref{fig:supp_NMR_orientation}(a)] and the comparison between the Knight shift and a squared-sine fit [\Figref{fig:supp_NMR_orientation}(b)]. 
\newpage

\subsection{Confirmation of bulk potassium stoichiometry}

\begin{figure}[ht!]
	\centering
	\includegraphics[width=\linewidth]{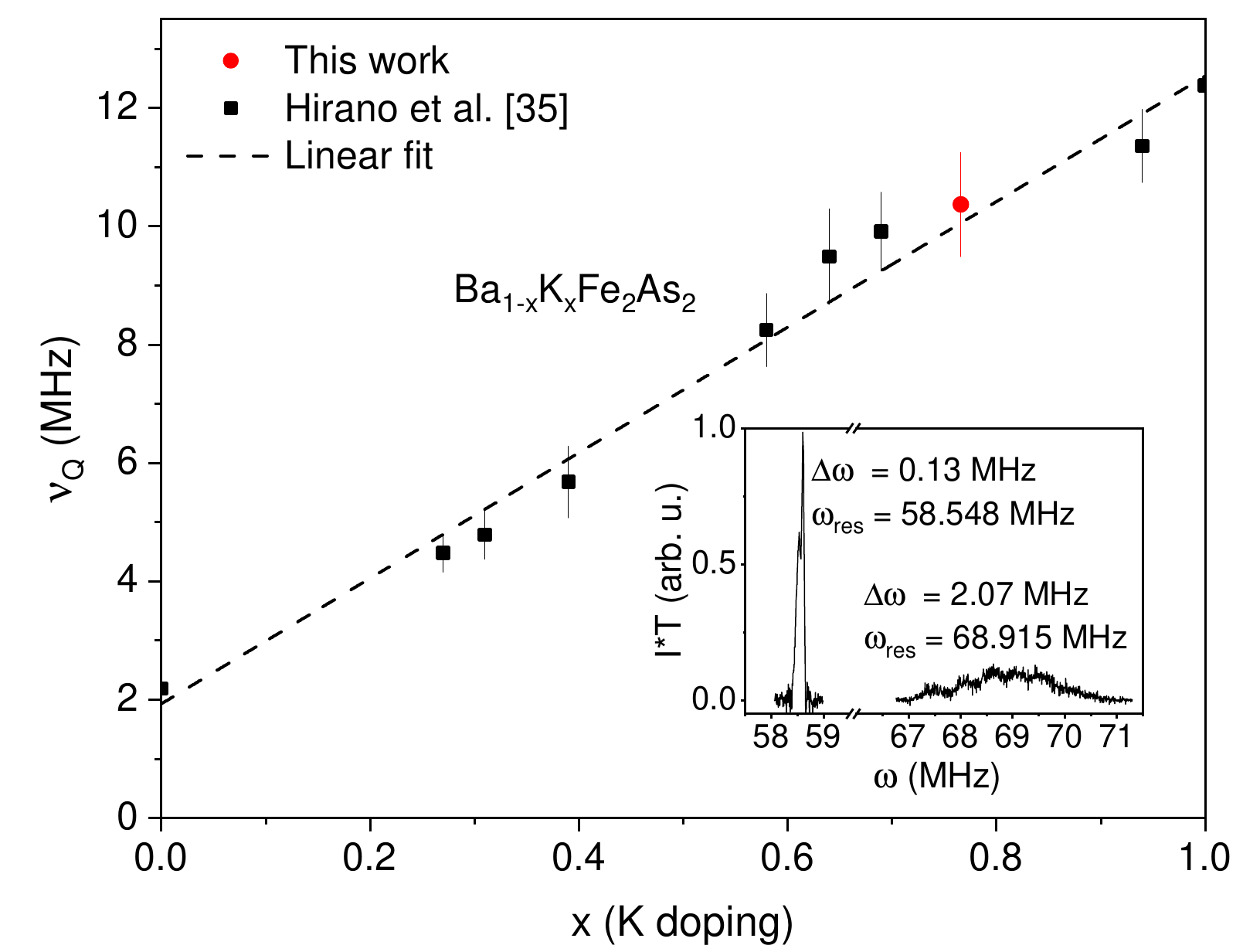}
	\caption{ Potassium-doping dependence of the quadrupole frequency $\nu_Q$ of \bkfa. Here, $\nu_Q = 10.367$ MHz is the quadrupole frequency of the $x=0.766(1)$ sample used for our NMR measurements (red circle), while the black rectangles represent data taken from Hirano \textit{et al.} \cite{hirano_potential_2012}. The dashed line is a linear fit to the data from Hirano \textit{et al}. The inset shows the central transition and the high-frequency satellite transition of the $^{75}$As spectrum of our NMR sample at 20 K and 8 T.}
	\label{fig:supp_quadrupole_frequency}
\end{figure}

In order to confirm the potassium doping level of our sample by means of NMR, we determined the quadrupole frequency $\nu_Q$ by measuring the central transition and the higher-frequency satellite transition at 20 K and 8 T (inset of \Figref{fig:supp_quadrupole_frequency}). Since, in the present case, the $c$ axis is the principle axis of the electric-field-gradient (EFG) tensor, the quadrupole frequency can be determined as the frequency difference between the first spectral moments of the central and satellite lines, yielding $\nu_Q = 10.367$ MHz. In the main panel of \Figref{fig:supp_quadrupole_frequency}, we compare this to the quadrupole frequencies measured by Hirano \textit{et al.}. Within the error bars, the quadrupole frequency of our sample fits very well into the systematic trend of these literature data. 

\newpage

\subsection{Superconducting transitions probed by RF penetration}

\begin{figure}[ht!]
	\centering
	\includegraphics[width=\linewidth]{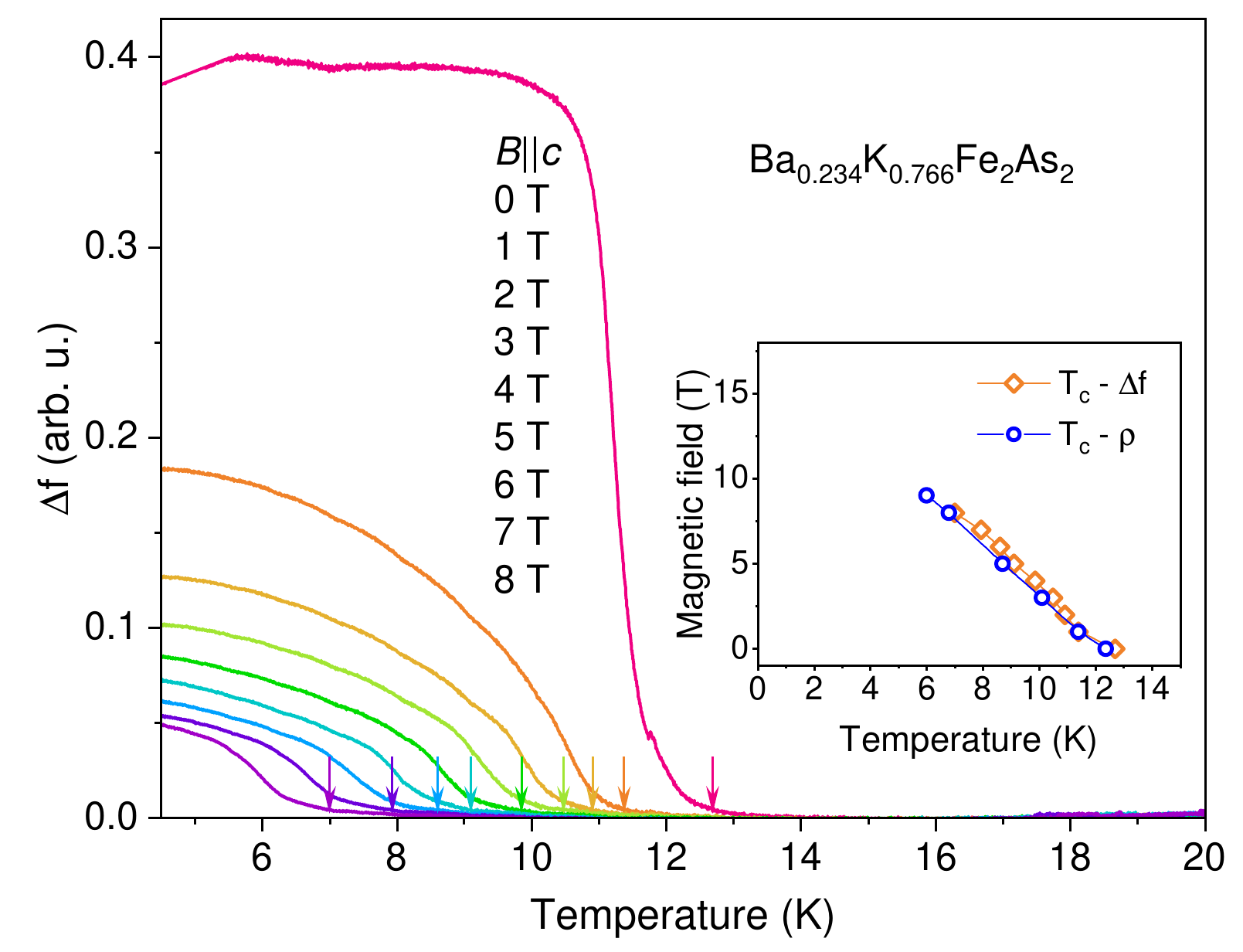}
	\caption{Temperature dependence of the resonance frequency shift $\Delta f$ at magnetic fields from 0 to 8 T. A linear background was subtracted from the raw data. A pronounced increase of $\Delta f$ occurs at $\Tc{}$, marked by the vertical arrows, and shifts towards lower temperatures with increasing magnetic-field strength. The inset compares $\Tc{}$ determined in this way (orange diamonds) with the $\Tc{}$ from resistivity measurements [blue circles, same data as shown in \Figref{fig:rxx_Nernst_SH}(d)].}
	\label{fig:supp_resonant_freq}
\end{figure}

As a confirmation of the superconducting phase diagram of our NMR sample as determined by means of the transport measurements presented in the main text, we also probed the shift $\Delta f$ of the resonance frequency $f_{res}$ of the NMR circuit at temperatures across the superconducting transition and magnetic fields applied parallel to the crystallographic $c$ axis. For this, we monitored changes of the complex radio-frequency (RF) reflection coefficient S11 at the NMR circuit, using a vector network analyzer. Here, the shift of the resonance frequency indicates a relative change of sample-volume penetration by the probing RF field. A linear background contribution of the measured data was subtracted, and the resulting curves of $\Delta f(T)$ are shown in \Figref{fig:supp_resonant_freq}. Here, $\Tc{}$ was determined as the temperature where $\Delta f$ is two orders of magnitude below the maximum value of the zero-field curve.

The inset of \Figref{fig:supp_resonant_freq} compares the superconducting transition temperatures $\Tc{}$ from the frequency-shift measurements with those determined by means of resistivity [\Figref{fig:rxx_Nernst_SH}(d)]. The results of both measurements agree very well, with a deviation of less than 0.5 K at each field. Such a small discrepancy is expected due to the different sensitivities of the contactless high-frequency technique and the electrical-transport measurements.

\newpage

\subsection{\texorpdfstring{$1/T_1T$}{1/T1} of \texorpdfstring{\bkfa}{Ba1-xKxFe2As2}}

\begin{figure}[ht!]
	\centering
	\includegraphics[width=\linewidth]{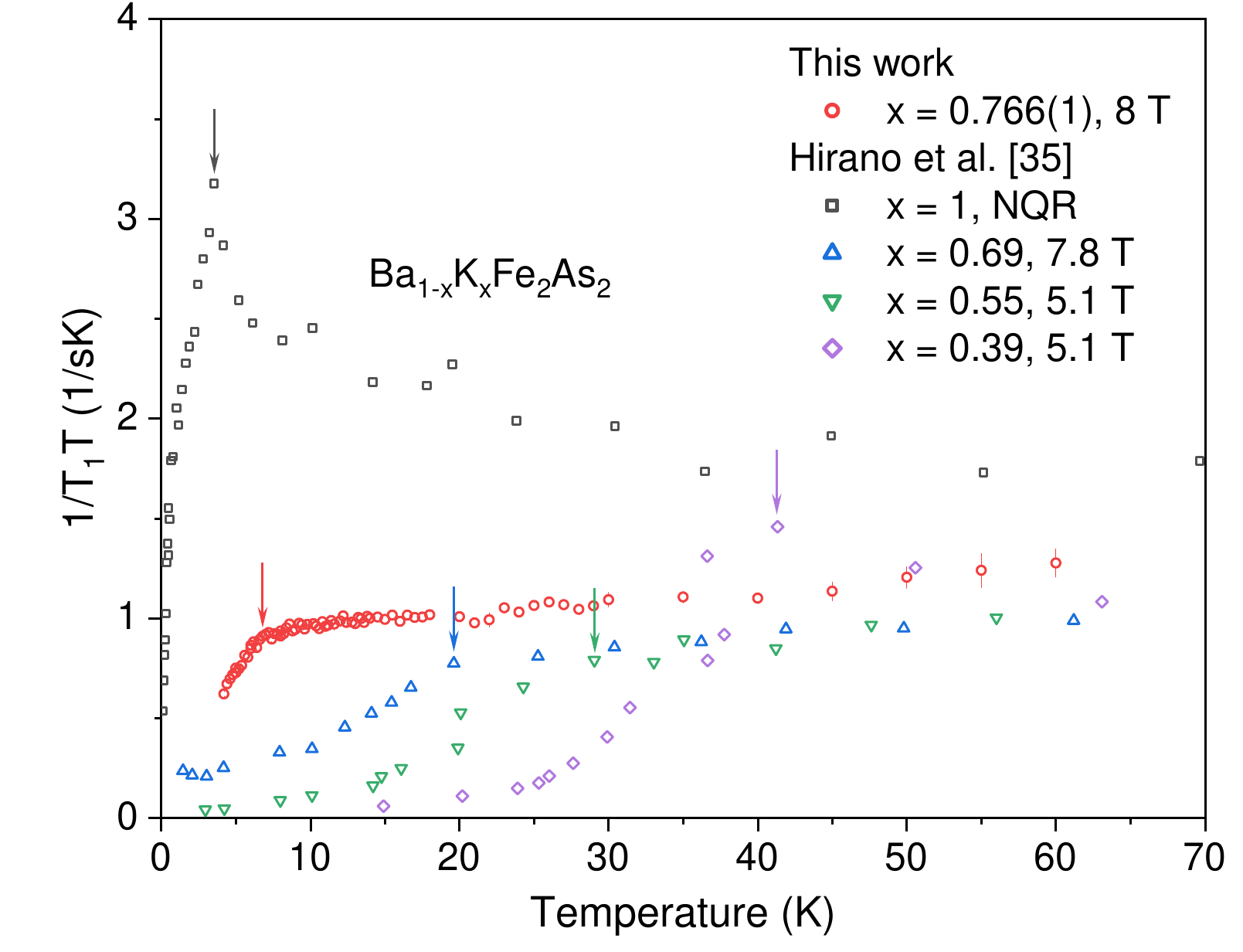}
	\caption{Spin-lattice relaxation rate $1/T_1T$ for $x=0.766(1)$ (red circles), compared to data for $x = 1,~0.69,~0.55,$ and$~0.39$ (gray rectangles, blue triangles, green triangles, and purple diamonds, respectively) from Hirano \textit{et al.} \cite{hirano_potential_2012}. The colored arrows mark $\Tc{}$ for the corresponding substitution levels.}
	\label{fig:supp_spin-lattice_relaxation_rate_Hirano}
\end{figure}

In \Figref{fig:supp_spin-lattice_relaxation_rate_Hirano}, we compare the $1/T_1T$ data of our NMR sample [$x=0.766(1)$] with results for \bkfa{} with different potassium contents from Hirano \textit{et al.} \cite{hirano_potential_2012}. Clearly, the data for our sample fit very well into the systematic behavior of these data from literature.

\newpage

\subsection{Gapped behavior at high temperatures}

As mentioned in the main text, the temperature-dependent $1/T_1T$ shows a monotonous increase with increasing temperatures above 25 K. A similar behavior has been reported, for example, for LaFeAsO$_{0.89}$F$_{0.11}$ by Ishida \textit{et al.} \cite{ishida_nmr_2011}. This dependence may be described by a gapped behavior, following an exponential form $1/T_1T \propto \exp(-\Delta_{PG}/T) + \text{const}$. 

\begin{figure}[ht!]
	\centering
	\includegraphics[width=\linewidth]{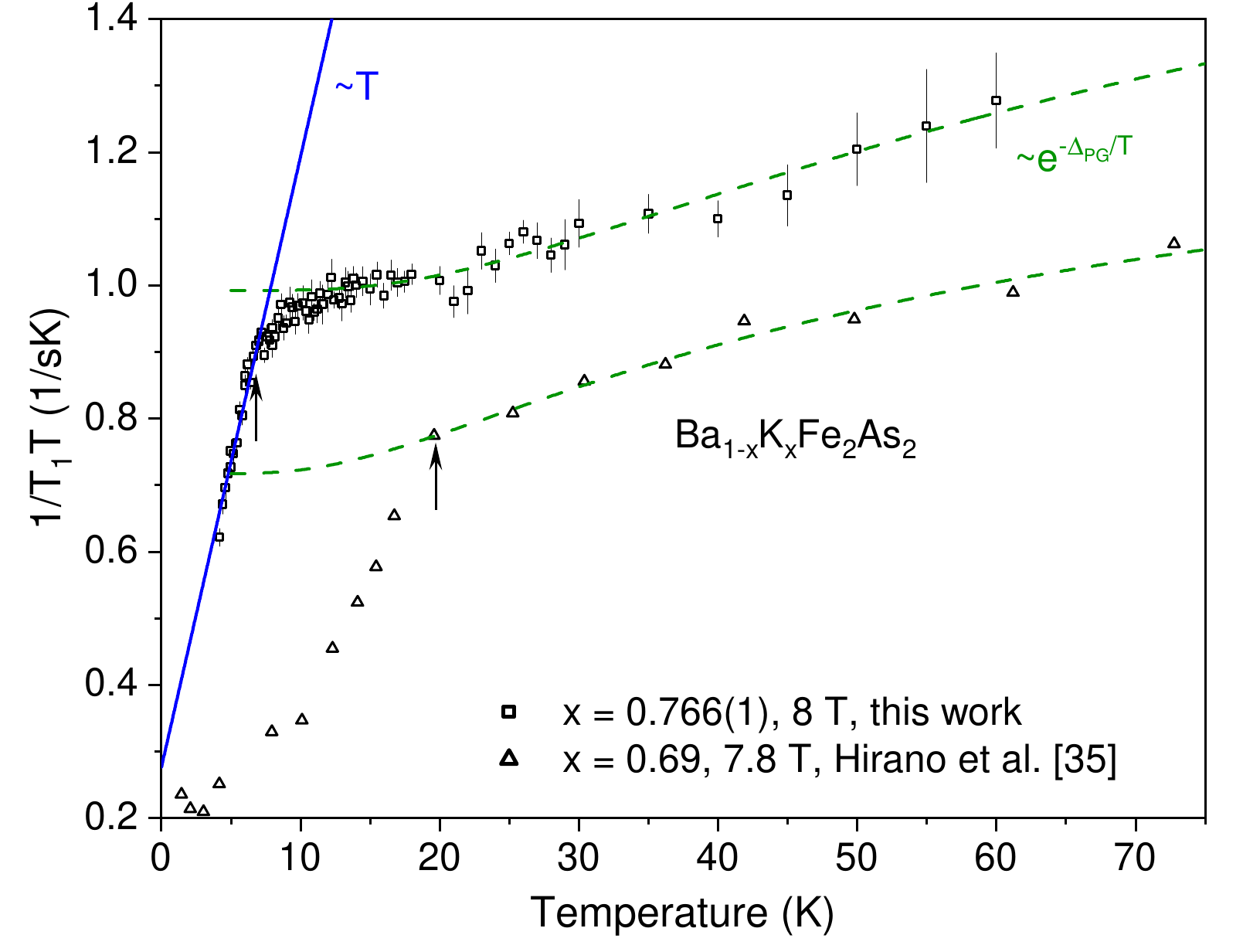}
	\caption{Temperature-dependent $1/T_1T$ data for \bkfa{} samples with $x=0.766(1)$ (squares) and $x=0.69$ (triangles). The latter data are taken from Hirano \textit{et al.} \cite{hirano_potential_2012}. The black arrows denote $\Tc{}$. The data at high temperatures are compared to an exponential fit (green dashed lines), describing a gapped behavior with a gap of approximately 76 K for our sample with $x=0.766(1)$. Here, a sharp kink at $\Tc{}=6.8$ K marks the transition to the superconducting state, with an almost linear low-temperature behavior (blue line). In the temperature range between 6.8 and 12 K, $1/T_1T$ shows a drop already above $\Tc{}$, indicating the opening of a pseudogap. The data for the $x=0.69$ sample are also compared to an exponential fit at high temperatures, giving a gap of 47 K.}
	\label{fig:supp_high_T_pseudogap_8T}
\end{figure}

In \Figref{fig:supp_high_T_pseudogap_8T}, we compare the regime above 12 K with the previously reported data of a $x=0.69$ sample from Hirano \textit{et al.} \cite{hirano_potential_2012}, and describe both with the gap equation. The gapped behavior provides a good description of both datasets in the high-temperature regime, yielding $\Delta_{PG}=76$ K for $x=0.766(1)$ and $\Delta_{PG}=47$ K for $x=0.69$. Importantly, the observed drop of $1/T_1T$ in the intermediate temperature regime $\Tc{} < T < T^*$ for $x=0.766(1)$ does not appear for the $x=0.69$ sample.

\newpage

\subsection{\texorpdfstring{$1/T_1T$}{1/T1} and specific-heat data at 16 T}

\begin{figure}[ht!]
	\centering
	\includegraphics[width=\linewidth]{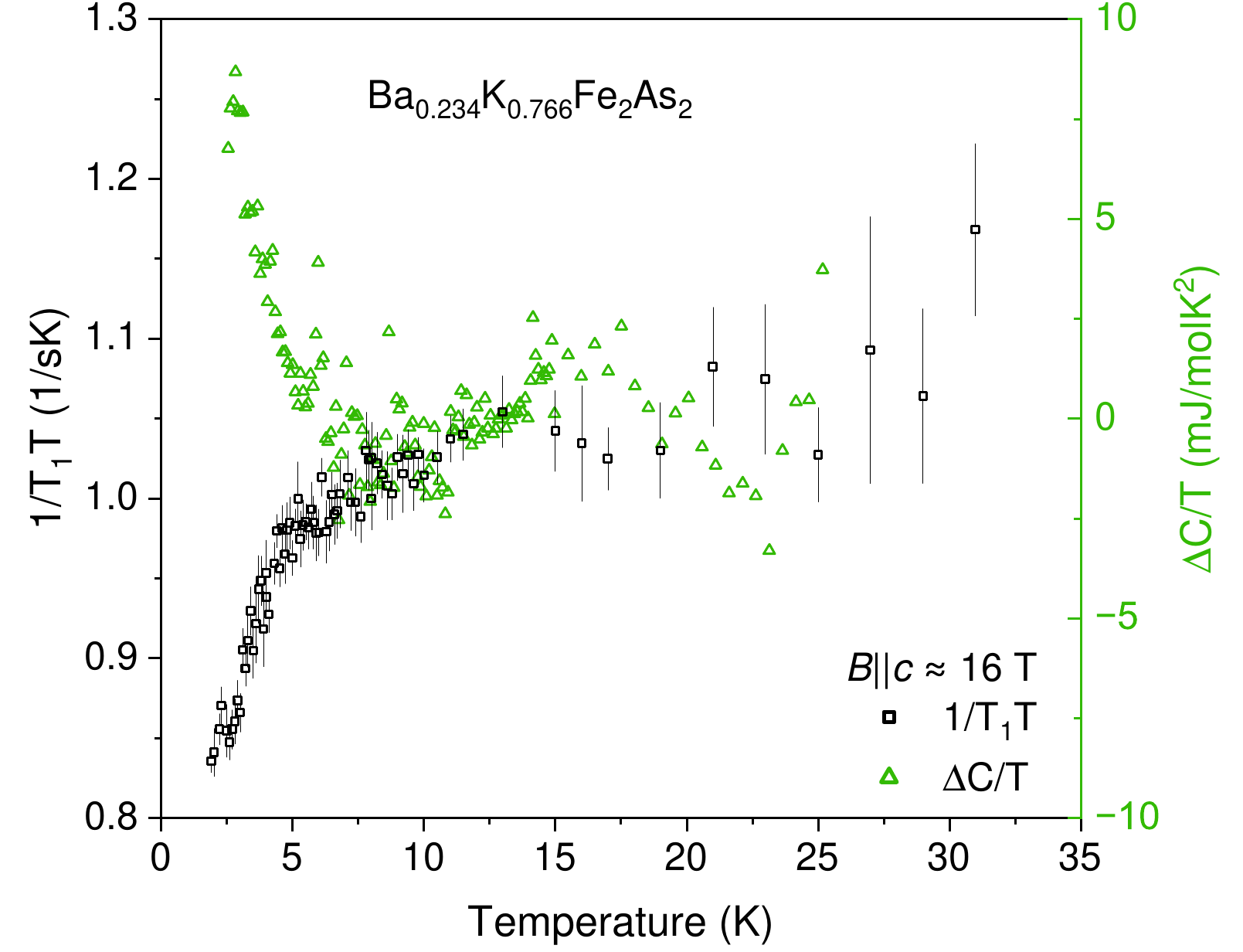}
	\caption{Temperature dependence of the spin-lattice relaxation rate (black rectangles) and specific-heat difference $\Delta C/T$ (green circles) at 16 T.}
	\label{fig:supp_spin-lattice_relaxation_rate_16T}
\end{figure}

In \Figref{fig:supp_spin-lattice_relaxation_rate_16T}, we show the temperature dependence of $1/T_1T$ for a field of 16 T and compare it to our specific-heat data recorded at the same field. The $1/T_1T$ data at 16 T reveal a similar pseudogap behavior as observed at 8 T, but at a slightly lower temperature of about 10.5 K, which also coincides with the onset temperature of the specific-heat increase at the same field. 
\newpage

\section{\texorpdfstring{$\mu$SR}{muSR}}
\begin{figure}[ht!]
	\centering
	\includegraphics[width=\linewidth]{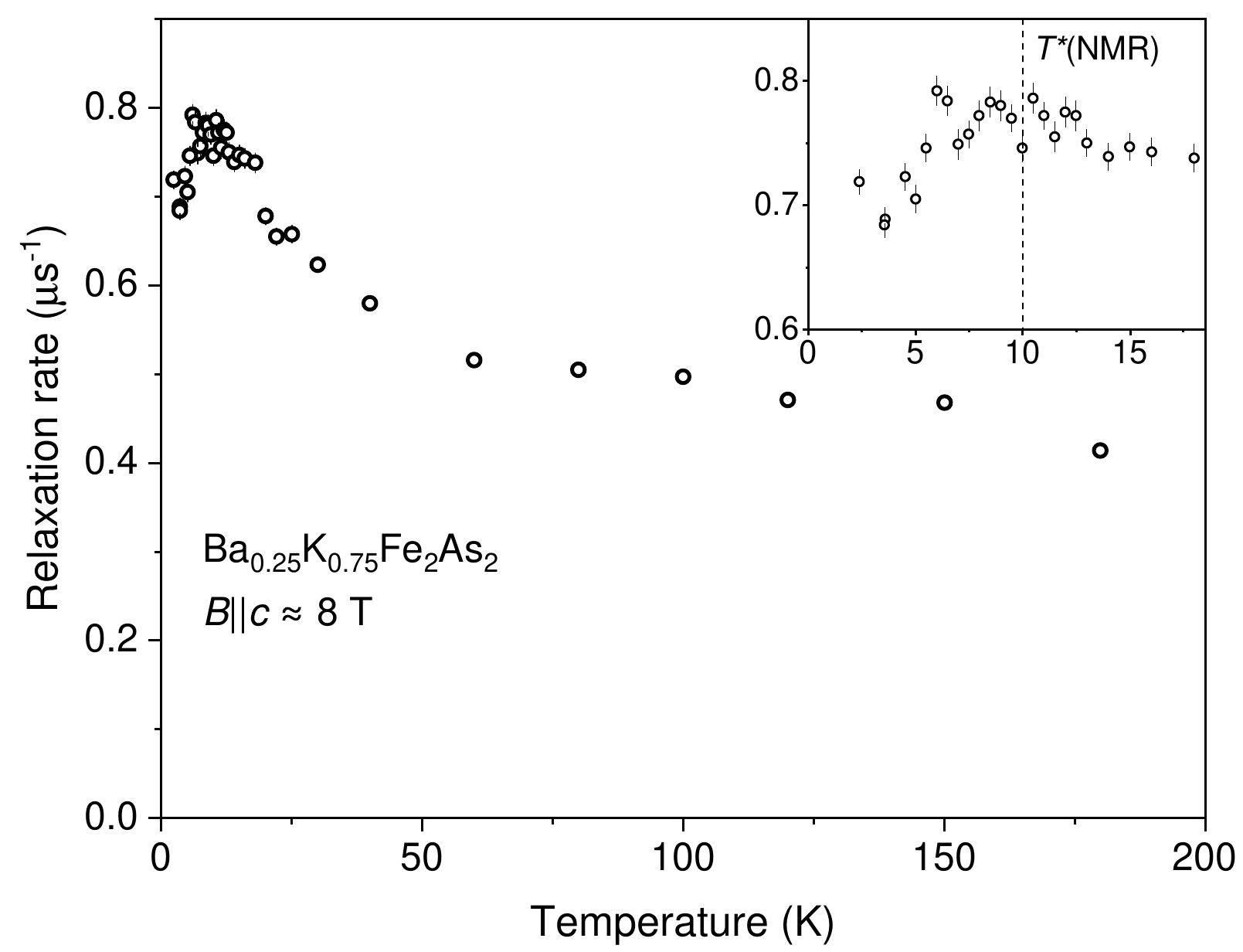}
	\caption{Temperature dependence of the transversal-field muon spin relaxation rate at 8 T obtained from the spectra shown in Fig.~\ref{fig:muSR}. The inset shows an enlarged view of the data at low temperatures.}
	\label{fig:muSR_sup}
\end{figure}

The temperature dependence of the muon-spin relaxation rate $\lambda$ is shown in Fig.~\ref{fig:muSR_sup}. It was obtained from the fit of the TF-$\mu$SR data shown in Fig.~\ref{fig:muSR} and corresponds to signal \RNum{1}. The TF-$\mu$SR line width is composed of both static and dynamic contributions. The increase of $\lambda$ below 50 K is related to static contributions of the individual crystals in the stack. Here, the variation of temperature dependencies of the Knight shift results in an enhancement, which is not seen in the spin-lattice relaxation rate (see Fig.~\ref{fig:NMR_spin-lattice_relaxation_rate_8T}). Despite this inhomogeneous broadening, $\lambda$ clearly reduces below $T^*$, consistent with results from other probes.

\newpage

\section{Scanning-tunneling-microscopy results}

\begin{figure}[ht!]
	\centering
	\includegraphics[width=\linewidth]{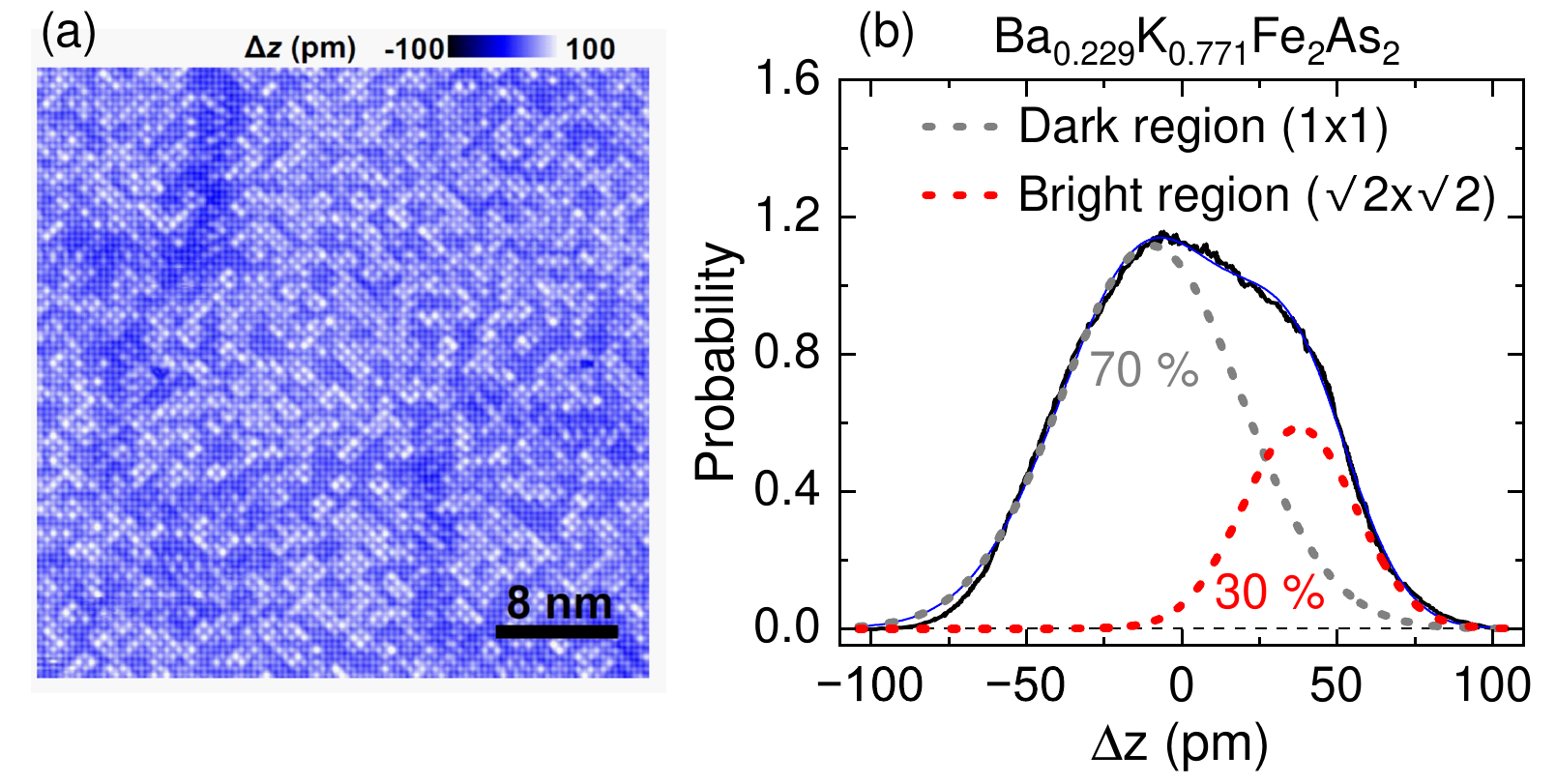}
	\caption{(a) Atomically-resolved topographic image taken from an As-terminated surface of Ba$_{0.229}$K$_{0.771}$Fe$_2$As$_2$ ($V = -10 \,\,\mathrm{mV}$, $I = 300 \,\mathrm{pA}$, image size: $40\times 40\,\mathrm{nm}^2$). The imaged area consists of regions characterized by $(1\times 1)$ unit cell of the As-lattice (dark regions) and those exhibiting ($\sqrt{2} \times \sqrt{2}$) spatial periodicity originating from surface reconstruction (bright regions). (b), Histogram showing the height distribution of the As-lattice points in (b).}
	\label{fig:STM}
\end{figure}

The STM topography image [Fig.~\ref{fig:STM}(a)] of the sample with $x = 0.771(1)$ shows that the As-terminated surface comprises randomly distributed bright and dark regions that exhibit different symmetries with respect to the ($1\times 1$) As lattice. Since the bright regions with the additional ($\sqrt{2}\times \sqrt{2}$) spatial order have different heights compared to the dark regions, a histogram of the measured heights at the lattice points exhibits two peaks, with the lower and higher peaks corresponding to dark and bright regions, respectively. The portion of different regions can be estimated by measuring the areas of these two peaks. Such a histogram is shown in Fig.~\ref{fig:STM}(b), where the black curve represents the experimental data. The gray and red dashed lines are the fitted results accounting for the dark and bright regions, respectively. The blue curve is the sum of the two. Further analysis reveals that the ratio between the areas under the gray and red curves is 7:3, a value close to that between the K and Ba atoms present in the material. The presence of regions of two different types can be attributed to an inhomogeneous substitution of Ba by K atoms.
\newpage
\section{\texorpdfstring{$T^*$}{T*} in transport measurements}
\begin{figure}[ht!]
	\centering
	\includegraphics[width=\linewidth]{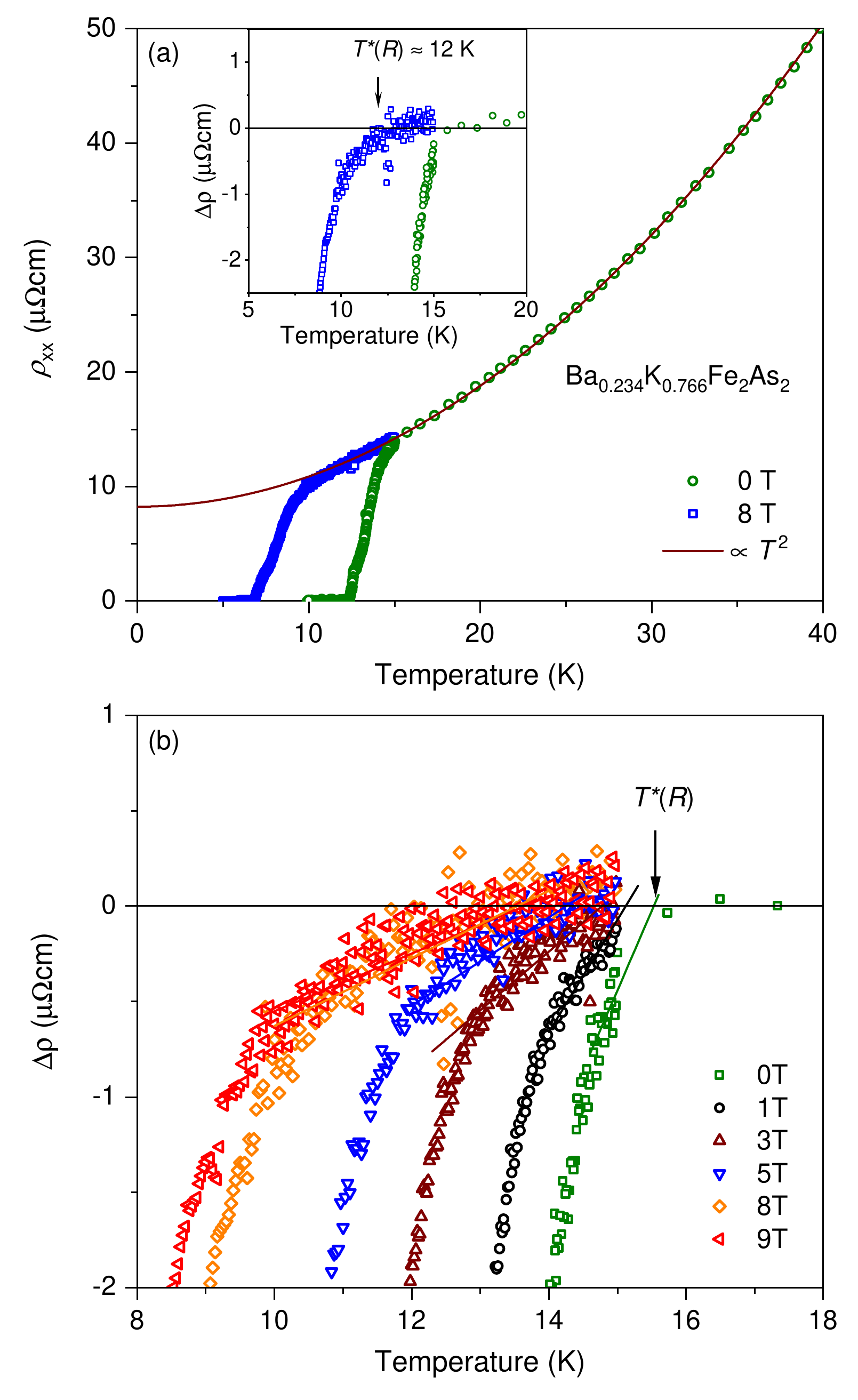}
	\caption{(a) Temperature dependence of the electrical resistivity in zero field and at 8 T. The resistivity up to 40 K follows a $T^2$ behavior. The inset shows the difference $\Delta \rho$ between the experimental data and the $T^2$ fit (solid line) near $T_c$. The resistivity deviates from the normal-state behavior at $T^*(R)$ slightly above $T^*(NMR)$ shown in Fig.~\ref{fig:NMR_spin-lattice_relaxation_rate_8T}. (b) Temperature dependence of $\Delta \rho$ around $\Tc{}$. The obtained $T^*(R)$ values are summarized in Fig.~\ref{fig:rxx_Nernst_SH}(d)}
	\label{fig:rho_pseudogap}
\end{figure}

The electrical resistivity of the NMR sample was measured by the 4-probe method. The temperature dependence in the normal state with $RRR = R(300K)/R_0\sim 70$ follows Fermi-liquid $T^2$ behavior below 40 K, indicating the high quality of the sample (see Fig.~\ref{fig:rho_pseudogap}(a)). This also indicates that the nanoscale inhomogeneities, observed in the STM measurements, do not noticeably affect electronic transport properties.

The electrical resistivity deviates from the $T^2$ behavior below $T^*$ (see Fig.~\ref{fig:rho_pseudogap}(b)). The obtained $T^*$ values from the transport measurements are summarized in Fig.~\ref{fig:rxx_Nernst_SH}(d), which within the error bars are consistent with the $T^*$ value obtained from the NMR.

\newpage

\section{Theoretical analysis  of the external-field-driven suppression of the critical temperatures $\Tc{\groupZ{2}}$  and  $T_c $ }

Simulations in an external field are performed with a Ginzburg-Landau theory derived from quasiclassical theory. Although, we believe that the effect is relatively generic for $s+is$ pairing symmetry and similar models. As an example, we used a minimal microscopic model of a clean superconductor with three overlapping bands at the Fermi level, which has been suggested to describe the superconducting state that breaks the time-reversal symmetry in the hole-doped 122 iron-pnictide compounds \cite{Stanev2010three,Maiti2013,Marciani.Fanfarillo.ea-2013}. 

\begin{figure}[!htb]
	\hbox to \linewidth{\hss
		\includegraphics[width=.9\linewidth]{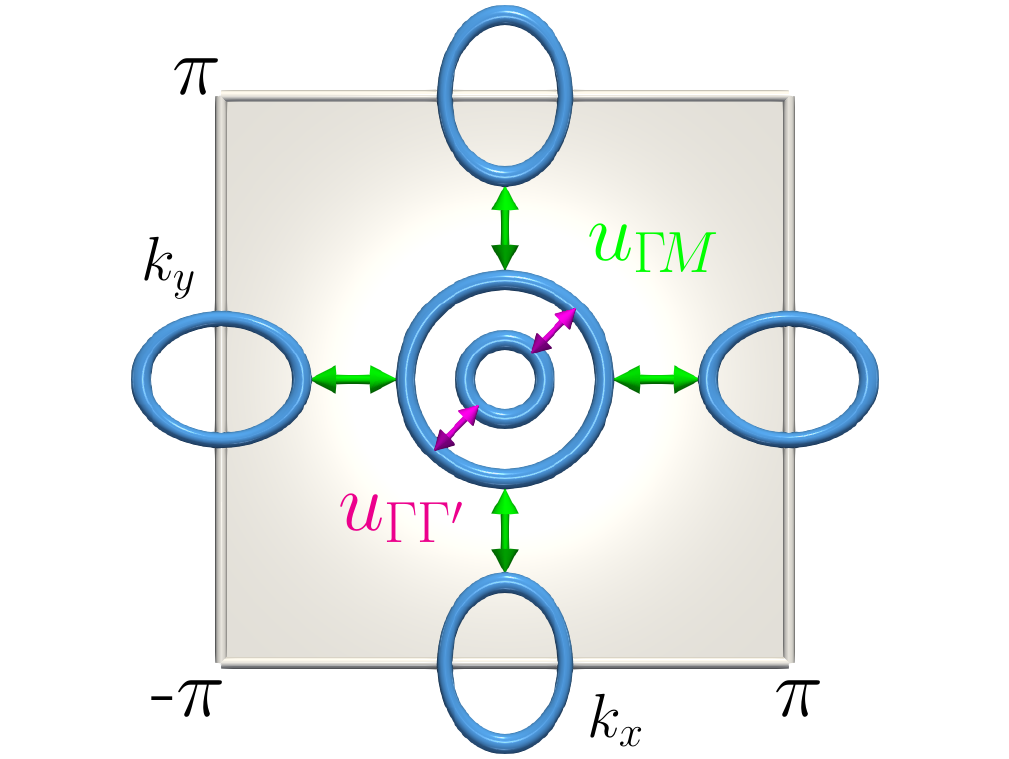}
		\hss}
	\caption{
		Schematic view of the gap structure in the first Brillouin zone reproducing certain features of the hole-doped iron pnictide compound  \bkfa. It consists of two Fermi-surface pockets at the $\Gamma$ point shown by circles and two other pockets at the $M$ points $(0; \pi)$ and $(\pi; 0)$ displayed by ellipses. As discussed in the text, within this model the $\sis$ state is favoured by the superconducting coupling that is dominated by the interband repulsion between $u_{\Gamma\!M}$ the $M$ and $\Gamma$/$\Gamma^\prime$ Fermi surfaces, as well as between both $\Gamma$/$\Gamma^\prime$ pockets $u_{\Gamma\Gamma^\prime}$.
	}
	\label{Fig:Micro-BZ}
\end{figure}

The model band structure illustrated in \Figref{Fig:Micro-BZ} consists of two Fermi-surface pockets at the $\Gamma$ point associated with gaps $\Delta_{1,2}$. A third gap $\Delta_3$ is associated with the Fermi-surface pocket at the $M$ point. The assumption here is that the crystalline $C_4$ symmetry is not broken and corresponds to an $s$-wave state. In the model considered here, the dominating pairing channels are chosen to be dominated by interband repulsion between the three bands. Such a scenario was suggested to be relevant for iron-based superconductors \cite{Maiti2013}. However, we stress that the nature of interaction and the number of gaps and their values are not fully known at the moment. We  expect that this assumption of pairing interaction is not very important and other multiband pairing constants also have a parameter space leading to similar effects. Note also that the $C_4$ symmetry of the model is not essential for the simulations and, at the moment, other similar states such as $s+id$ are not completely ruled out \cite{halcrow2024probing}.

\subsection{Microscopic derivations of the model}

Within the quasiclassical approximation, the model depends on various parameters of each Fermi surface (the partial densities of states $\nu_a$ and the Fermi velocities ${\bs v}^{(a)}_{F}$), where the index $a=1,2,3$ labels the different bands. While the conclusions should be rather generic for a broad class of the models, as an example,  we consider the case of an interband-dominated repulsive pairing, which was proposed to be relevant for iron-based superconductors \cite{Maiti2013}. There, the pairing, which may lead to the broken time-reversal symmetry because, of the competing interband repulsive channels is, described by the coupling matrix: 
\Equation{EqSM:CouplingMatrix}{
	\hat\Lambda =  \left(%
	\begin{array}{ccc}
		0        & -u_{\Gamma\Gamma^\prime}    & -u_{\Gamma\!M}   \\
		-u_{\Gamma\Gamma^\prime}     & 0       & -u_{\Gamma\!M}   \\
		-u_{\Gamma\!M}  & -u_{\Gamma\!M} & 0        \\
	\end{array}  \right) \,.
}
The coefficients $u_{\Gamma\Gamma^\prime}$ and $u_{\Gamma\!M}$ describe the interactions between the two Fermi surfaces around the $\Gamma$ point and between the Fermi surfaces around the $\Gamma$ point with those at the $M$ point, respectively.

Expressing the solutions of the Eilenberger equations for the quasiclassical propagators, as power expansions of the gap-function amplitudes $\Delta_{a}$ and of their gradients yields the Ginzburg-Landau equations (see details in \cite{Garaud2017microscopically,Garaud2022effective}):
\Equation{EqSM:GL}{
	\big[(G_0+\tau-\hat\Lambda^{-1}){\bs\Delta} \big]_a
	= -K^{(a)}_{ij}D_iD_j\Delta_a + |\Delta_a|^2\Delta_a  \,,
}
where ${\bs \Delta} = (\Delta_1,\Delta_2,\Delta_3)^T$. Here, $K^{(a)}_{ij} =\hbar^2\rho\expval{v^{(a)}_{Fi}v^{(a)}_{Fj}}_a /2\Tc{}^2$ is the anisotropy tensor, where the indices $i,j$ stand for the $x,y$ coordinates and the average is taken over the $a$-th Fermi surface.

The critical temperature is determined by the smallest positive eigenvalue of the inverse coupling matrix $\hat\Lambda^{-1}$. That is, if $G_n$ denote the positive eigenvalues of the inverse coupling matrix $\hat\Lambda^{-1}$, the critical temperature is determined by the equation $G_0 = \min_n (G_n)$. Provided that all eigenvalues are positive, the number of components of the effective field theory coincides with the number of bands. 

For an interband-dominated repulsive pairing, the inverse coupling matrix $\hat\Lambda^{-1}$ has only two positive eigenvalues. Fields that can nucleate are those associated with positive eigenvalues and, therefore, the resulting Ginzburg-Landau theory that we will use has two components. Thus, the three gaps $\Delta_a$ are given as linear combinations of only two complex fields $\psi_{1,2}$, and the eigenvectors ${\bs\Delta}_a$ associated with positive eigenvalues,  by expressing the general order parameter as 
\Align{EqSM:OProtation}{
	{\bs \Delta}&=\psi_1{\bs \Delta}_1+\psi_2{\bs \Delta}_2\,, \nonumber \\
	\text{and}~~~(\Delta_1,\Delta_2,\Delta_3)&=(x\psi_2-\psi_1,x\psi_2+\psi_1,\psi_2)\,.
}
Here, the positive eigenvalues of $\hat\Lambda^{-1}$ explicitly reads as $G_1 = 1/u_{\Gamma\Gamma^\prime}$ and $G_2 =\left(u_{\Gamma\Gamma^\prime}+\sqrt{u_{\Gamma\Gamma^\prime}^2+8u_{\Gamma\!M}^2} \right) /4u_{\Gamma\!M}^2$. The associated eigenvectors are ${\bs\Delta}_1=(-1,1,0)^T$ and ${\bs\Delta}_2=(x,x,1)^T$, where $x=(u_{\Gamma\Gamma^\prime}-\sqrt{u_{\Gamma\Gamma^\prime}^2+8u_{\Gamma\!M}^2})/4u_{\Gamma\!M}$. The substitution of the linear combination \Eqref{EqSM:OProtation} into the Ginzburg-Landau Eqs. \Eqref{EqSM:GL}, after projection onto the eigenvectors ${\bs \Delta}_{1,2}$, yields a system of two Ginzburg-Landau equations for the fields $\psi_{1,2}$ \cite{Garaud2017microscopically}. For discussions about the relevance of reducing to a two-component model, see \cite{Yerin2013,Aase.Johnsen.ea-2023}.

The two-component free-energy functional that corresponds to the Ginzburg-Landau Eqs. \Eqref{EqSM:GL}, and whose variations with respect to the vector potential $\A$ give the supercurrent, reads as (in dimensionless units): 
\Align{EqSM:FreeEnergy}{
	\F =\frac{\B^2}{2}+&\frac{1}{2}\sum_{a,b=1}^2
	k_{ab}(\D\psi_a)^*\D\psi_b
	+V({\bs\Psi}) \,,	
}

\Align{EqSM:FreeEnergyV}{\text{where}~V({\bs\Psi}) &= \sum_{a,b=1}^2a_{ab}\psi_a^*\psi_b
	+\frac{b_{ab}}{2}|\psi_a|^2|\psi_b|^2 
	\\
	&~~~+\frac{c_{12}}{2}\big(\psi_1^{*2}\psi_2^2+c.c.\big) 
	\,.\nonumber
}
Here, the complex fields $\psi_a=|\psi_a|\Exp{i\varphi_a}$ are the $a={1,2}$ components of the superconducting order parameter $\Psi^\dagger=(\psi_1^*,\psi_2^*)$. They are electromagnetically coupled by the vector potential $\A$ of the magnetic field $\B=\Curl\A$, through the gauge derivative $\D\equiv\Grad+ie\A$. There, the coupling constant $e$ is used to parametrize the London penetration length.

The parameters of the potential term of the Ginzburg-Landau free energy \Eqref{EqSM:FreeEnergy} are expressed, in terms of the coefficients of the coupling matrix \Eqref{EqSM:CouplingMatrix} as 
\Align{EqSM:Parameters}{
	a_{jj} &= -|\bs \Delta_j|^2(G_0-G_{j}+\tau )\,,~~a_{12}=0 \\
	b_{11} &= 2    \,,~~b_{22}=(2x^4+1) \,,~~ b_{12} = 4x^2 \,,~c_{12}=2x^2 
	\,, \nonumber
}
where $|\bs \Delta_1|^2 =2 $ and $|\bs \Delta_2|^2=2x^2+1$. The $s+is$ state is 
symmetric under $C_4$ transformations, thus, the coefficients satisfy 
$K^{(j)}_{xx} = K^{(j)}_{yy} = K^{(j)}$. As a results, the coefficients of the 
gradient terms in \Eqref{EqSM:FreeEnergy} read as 
\SubAlign{EqSM:Parameters:K}{
	k_{11} & = 2\xi_0^{-2}\big[ K^{(1)} +  K^{(2)} \big]				\\
	k_{22} & = 2\xi_0^{-2}\big[(K^{(1)} + K^{(2)})x^2 + K^{(3)} \big]	\\
	k_{12} =k_{21} & = 2\xi_0^{-2} x\big[K^{(2)} - K^{(1)} \big] \,.
}

Note that for the energy to be positive definite, the coefficients of the kinetic terms 
should satisfy the relation $k_{11}k_{22}-k_{12}^2>0$. Also, for the free-
energy functional to be bounded from below, the coefficients of the terms that are fourth 
order in the condensates should satisfy the condition $b_{11}b_{22}-(b_{12}+c_{12})^2>0$. 
Finally, the condition for having a nonzero ground-state density is 
$\det\hat{a}=a_{11}a_{22}-a_{12}^2<0$. These conditions are, of course, satisfied by 
the microscopically calculated value \Eqref{EqSM:Parameters} and 
\Eqref{EqSM:Parameters:K}. Examples of their values used for numerical simulations are displayed in Table \ref{TabSM:GL-coefficents}.

\newcolumntype{P}[1]{>{\centering\arraybackslash}p{#1}}
\newcommand{\COLSIZ}{0.0675}
\begin{table*}[!ht]
	\caption{
		Coefficients of the Ginzburg-Landau free energy functional \Eqref{EqSM:FreeEnergy}, that correspond to the numerical simulations reported in \Figref{Fig:Field-heated}. The coefficients are calculated using the formulas \Eqref{EqSM:Parameters:K} and \Eqref{EqSM:Parameters}, derived consistently from the microscopic quasiclassical theory \cite{Garaud2017microscopically}. Other parameters of the microscopic model are $u_{\Gamma\!M}=0.5$, $u_{\Gamma\Gamma^\prime}=0.4875$, $K^{(1)}=1.0$, $K^{(3)}=0.015$, $e=0.5$, and $B_0=0.6$ for both simulations.
	}
	\centering
	\begin{tabular} 
		{|P{0.1\linewidth}|P{0.055\linewidth}|
			P{\COLSIZ\linewidth}P{\COLSIZ\linewidth}P{\COLSIZ\linewidth}|
			P{\COLSIZ\linewidth}P{0.05\linewidth}P{\COLSIZ\linewidth}|
			P{\COLSIZ\linewidth}P{\COLSIZ\linewidth}P{\COLSIZ\linewidth}|
			P{\COLSIZ\linewidth}|
		}\hline 
		
		Simul. 	& $T/\Tc{\mathrm{MF}}$	& $k_{11}$ 	& $k_{12}$ & $k_{22}$
		& $a_{11}$ 	& $a_{12}$ & $a_{22}$	
		& $b_{11}$ 	& $b_{12}$ & $b_{22}$	
		& $c_{12}$	\\
		
		\hline  \hline 
		&	0.90	
		& 2.2000	& 0.9075  & 0.5893
		& -0.1668	& 0       & -0.1508
		& 2.0000	& 1.0168  & 1.1292
		& 0.5084				\\
		$\mbf{(a)}$     &	0.80		
		& 2.2000	& 0.9075  & 0.5893
		& -0.3668	& 0       & -0.3017
		& 2.0000	& 1.0168  & 1.1292
		& 0.5084				\\
		$K^{(2)}=0.10$  &	0.70
		& 2.2000	& 0.9075  & 0.5893
		& -0.5668	& 0       & -0.4525
		& 2.0000	& 1.0168  & 1.1292
		& 0.5084				\\
		&	0.60	
		& 2.2000	& 0.9075  & 0.5893
		& -0.7668	& 0       & -0.6034
		& 2.0000	& 1.0168  & 1.1292
		& 0.5084				\\
		\hline  
		&	0.90	
		& 3.9000	& 0.0504  & 1.0214
		& -0.1668	& 0       & -0.1508
		& 2.0000	& 1.0168  & 1.1292
		& 0.5084				\\
		$\mbf{(b)}$     &	0.80		
		& 3.9000	& 0.0504  & 1.0214
		& -0.3668	& 0       & -0.3017
		& 2.0000	& 1.0168  & 1.1292
		& 0.5084				\\
		$K^{(2)}=0.95$  &	0.70
		& 3.9000	& 0.0504  & 1.0214
		& -0.5668	& 0       & -0.4525
		& 2.0000	& 1.0168  & 1.1292
		& 0.5084				\\
		&	0.60	
		& 3.9000	& 0.0504  & 1.0214
		& -0.7668	& 0       & -0.6034
		& 2.0000	& 1.0168  & 1.1292
		& 0.5084				\\
		\hline \hline  
	\end{tabular}
	\label{TabSM:GL-coefficents}
\end{table*}

\subsection{Ground state and time-reversal symmetry}

The superconducting ground state is the state which minimizes the potential energy 
\Eqref{EqSM:FreeEnergyV} 
and that is constant in space: $\Psi_0:=\mathrm{argmin}\, V({\Psi})$, with $\Psi_0^\dagger\Psi_0=const.\neq0$. The breakdown of time-reversal symmetry typically occurs due to a competition between different phase-locking terms, between the different condensates. Heuristically, the time-reversal symmetry of the ground-state can be understood as the invariance of $\Psi_0$ under complex conjugation (up to global $\groupU{1}$ transformations). Conversely, a superconducting state that spontaneously breaks time-reversal symmetry satisfies the condition
\Equation{EqSM:BTRS:Condition}{
	\Psi_0^*\Exp{i\chi}\neq\Psi_0~~\forall\chi\,.
}
For example, in the case of two components:
\Align*{
	\Psi_0&=(1,- 1)/\sqrt{2}:~~~\,\Psi_0^*=(1,- 1)/\sqrt{2}=   \Psi_0	\nonumber\\
	&~~~~~
	\Longrightarrow  \text{TRS\phantom{B}}		\nonumber \\
	\Psi_0^\pm&=(1,\pm i)/\sqrt{2}:~~~\Psi_0^{\pm*}=(1,\mp i)/\sqrt{2}=\Psi_0^\mp
	\nonumber \\
	&~~~~~\Psi_0^\mp
	\neq\Psi_0^\pm \Longrightarrow \text{BTRS}		\nonumber	\,.	 
}

The invariance under complex conjugation (related to the time-reversal operation) implies that the potential energy does not change if the sign of all relative phases $\varphi_{12}:=\varphi_{2}-\varphi_{1}$ is changed, $\varphi_{12}\to-\varphi_{12}$. It follows that if the relative phase $\varphi_{12}$ is not an integer multiple of $\pi$, then the ground state has a discrete $\groupZ{2}$ degeneracy in addition of the usual $\groupU{1}$ symmetry. This is the case of the potential \Eqref{EqSM:FreeEnergyV} 
studied here.

The potential term \Eqref{EqSM:FreeEnergyV}  
indicates that the ground-state relative phase will be $\pm\pi/2$ in the broken time-reversal symmetry phase. Hence, the component $n_y=2\Im(\psi_1^*\psi_2)/\Psi^\dagger\Psi$ is a good order parameter to measure the breakdown of time-reversal symmetry that can be viewed as easy-axis modification of the pseudospin $\bf n$. Examples of the ground state and simulations in external field are presented in \Figref{Fig:Field-heated-averages}.

Note that this order parameter is fourth-order in fermionic fields. As demonstrated by Monte-Carlo calculations \cite{Bojesen2013time,bojesen2014phase}, ordering of $\bf n$ can take place at temperatures higher than superconducting ordering (i.e. without long-range order in $\psi_i$). Hence, there  is no BCS gap in this resistive state, and pseudogap features are expected).

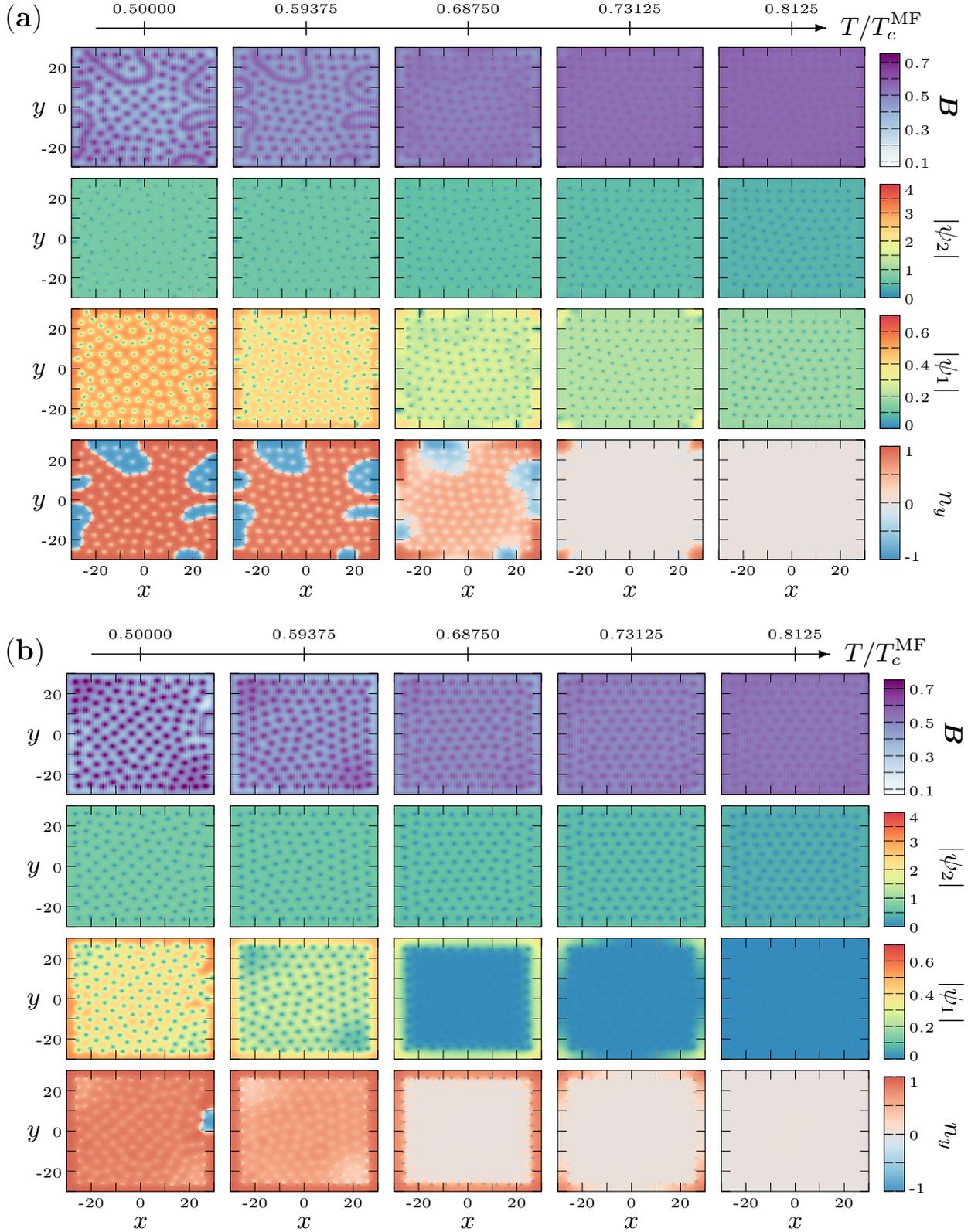
\begin{figure*}[htp]
	\hbox to \linewidth{\hss
		\includegraphics[width=.89\linewidth]{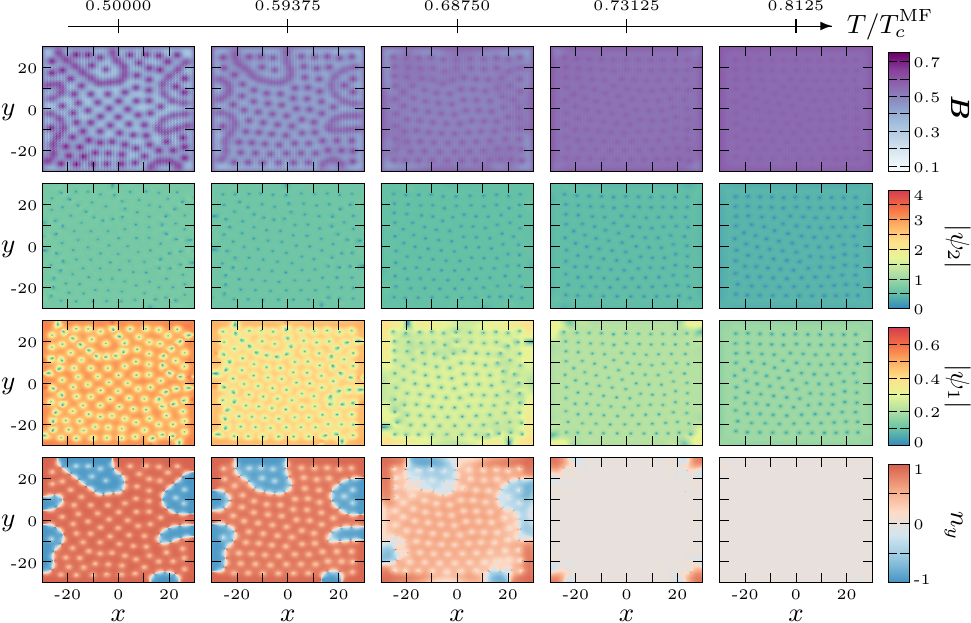}
		\hss}
	%
	\vspace{0.5cm}
	\hbox to \linewidth{\hss
		\includegraphics[width=.9\linewidth]{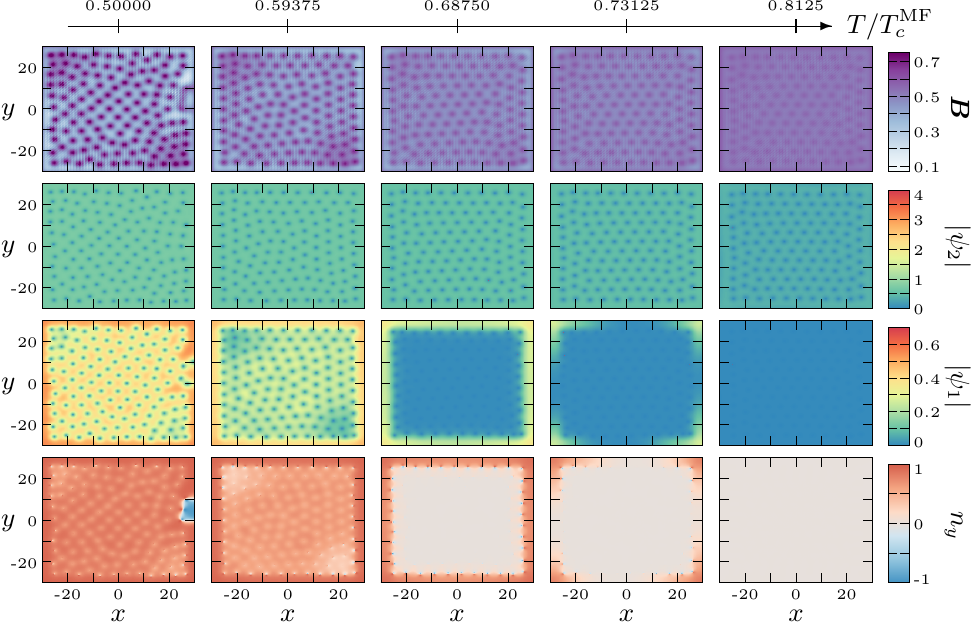}
		\hss}
	
	\begin{picture}(0,0)%
		\put(-240,600){\color{black}\Large $\mbf{(a)}$ }
		\put(-240,290){\color{black}\Large $\mbf{(b)}$ }
	\end{picture}
	\vspace{-0.25cm}
	\caption{
		Simulations of field-heated experiments. Note that increasing the temperature puts the system closer to $\Hc{2}$, in a similar way as increasing the applied field. The parameters of the microscopic model are $u_{\Gamma\!M}=0.5$, $u_{\Gamma\Gamma^\prime}=0.4875$, $K^{(1)}=1.0$, $K^{(3)}=0.015$, $e=0.5$, and $B_0=0.6$ for both simulations. The difference in is that in (a) $K^{(2)}=0.1$, while in (b) it is  $K^{(2)}=0.95$. The top row shows the magnetic field, the two middle rows show the densities $|\psi_a|$. The last line shows $n_y=2\Im(\psi_1^*\psi_2)/\Psi^\dagger\Psi$. Thus, since $n_y\neq0$, when the time-reversal symmetry is broken, the vector field ${\bs n}$ illustrates the restoration of the time-reversal symmetry when the system is approaching the upper critical magnetic field. 
	}
	\label{Fig:Field-heated}
\end{figure*}

\subsection{Simulations in an external field}

Superconductors in an external magnetic field $B_0{\bs{\hat z}}$ are characterized by the total (Gibbs) free energy ${G}=\int_\Omega\F-2\B\cdot B_0{\bs{\hat z}}$. In order to numerically investigate the properties in an external field, the vector potential $\A$ and the superconducting order parameter $\psi_{1,2}$ are discretized using a finite-element framework \cite{Hecht-2012}. The Ginzburg-Landau free energy is minimized using a non-linear conjugate gradient algorithm \cite{Nocedal.Wright}. For further details on the numerical methods employed here, see, for example, the related discussion in Ref. \cite{Garaud.Babaev.ea-2016}. 

\begin{figure}[!tb]
	\vspace{0.25cm}
	\hbox to \linewidth{\hss
		\includegraphics[width=\linewidth]{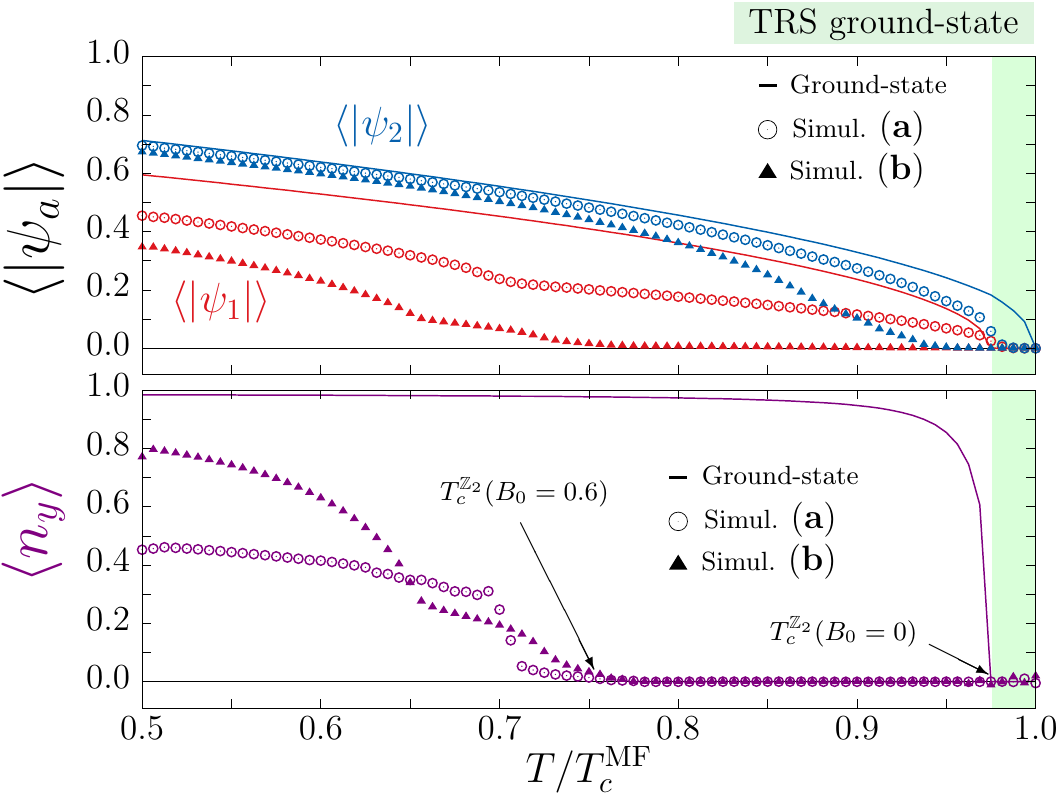}
		\hss}
	\caption{
		Average values of order parameters $\expval{\abs{\psi_a}}$ and $\expval{n_y}$ as a function of temperature compared for the ground state with simulations in an external field ($B_0=0.6$). Here, $n_y=2\Im(\psi_1^*\psi_2)/\Psi^\dagger\Psi$ is the order parameter that measures the spontaneous breakdown of the time-reversal symmetry. Note that it is a pseudospin vector that has no Zeeman-type coupling to the external magnetic field. Clearly, in contrast to the role of external magnetic fields in magnets, here the applied field suppresses the critical temperature of time-reversal symmetry breaking associated with the interband phase difference: $\Tc{\groupZ{2}}(B_0\neq0)\leq\Tc{\groupZ{2}}(B_0=0)$ (here, $\Tc{\groupZ{2}}(B_0=0)/\Tc{\mathrm{MF}}=0.975$).  
		Note that, since simulations deal with a finite sample, there are boundary effects. The cusp around $T/\Tc{\mathrm{MF}}=0.65$ for $\expval{n_y}$ in the simulation corresponds to a BTRS phase that survives near edges, while it is restored in the bulk. Irrespectively of that, the breaking of time-reversal symmetry is suppressed by an external field.
	}
	\label{Fig:Field-heated-averages}
\end{figure}

We performed several simulations of field-heated experiments of mesoscopic samples in a superconducting state, for different parameter sets. The Figs.~\ref{Fig:Field-heated-averages} and \ref{Fig:Field-heated} report global measured quantities, and the state of the system for different parameter sets denoted $\mbf{(a)}$ and $\mbf{(b)}$. In some sense, increasing the temperature drives the system closer to $\Hc{2}$, similarly than increasing the applied field. In the numerical simulations, vortex penetration in finite domains may be substantially affected by the Bean-Livingstone barrier. Hence, we first performed a \textit{field-cooled} simulation starting at a fixed external field above $\Tc{\mathrm{MF}}$, and then gradually decreased the temperature through both critical temperatures $\Tc{}$ and $\Tc{\groupZ{2}}$ to a low-temperature state, in which the time-reversal symmetry is broken. At each of the temperature steps, the Gibbs energy ${G}=\int_\Omega\F-2\B\cdot B_0{\bs{\hat z}}$ is minimized until convergence of the non-linear conjugate gradient algorithm is reached. The low-temperature states are then used as starting configurations for the \textit{field-heated} simulation, where the temperature is increased through both critical temperatures until destruction of superconductivity. In the cooling process, the system undergoes a second-order phase transition, at which the time-reversal symmetry is spontaneously broken at $\Tc{\groupZ{2}}$. 

Domain walls will naturally form during such a phase transition, where a discrete symmetry is broken. Production of domain walls during the different \textit{field-cooled} simulations strongly depends on details of the simulation and, hence, one may either end up on a low-temperature state with [\Figref{Fig:Field-heated}(a)] or without domain-walls as in [\Figref{Fig:Field-heated}(b)]. The microscopic parameters are given in the caption of \Figref{Fig:Field-heated}. The corresponding values of the Ginzburg-Landau parameters of the potential \Eqref{EqSM:Parameters} and kinetic terms \Eqref{EqSM:Parameters:K} are given in the Table \ref{TabSM:GL-coefficents}. Irrespectively of simulation parameters, the temperature of the spontaneous breaking of time-reversal symmetry is suppressed by an external field in this model.

Note that the Ginzburg-Landau-based analysis is not valid at low temperatures. Also, since it is based on mean-field approximation, it is not valid close to $\Tc{}$ due to strong fluctuation effects, hence, the calculations discussed below are restricted to temperatures below the critical temperature and show that $\Ztwo$ symmetry associated with two-fold degenerate phase locking can be restored by external field even below the superconducting critical temperature. Hence, for these parameters, weaker fields eliminate the quartic phase.

\subsection{Symmetry restoration at fields significantly larger than \texorpdfstring{$\Hc{1}$}{Hc1}}

The subdominant component $\psi_1$ is more suppressed by applied field than the dominant component $\psi_2$. This is clearly seen in Figs.~\ref{Fig:Field-heated-averages} and \ref{Fig:Field-heated}, in particular with parameter set $\mbf{(b)}$. Since one of the components disappears in the vicinity of $\Tc{\groupZ{2}}$, the phase-locking term, $\big(\psi_1^{*2}\psi_2^2+c.c.\big)$, in the free energy favors no particular relative phase, and the time-reversal symmetry is restored. This is the reason why the temperature at which the time-reversal symmetry is broken is lowered when an external field is applied.
Note that in an external field, for a different parameter set, the subtle interplay between the different components might produce the opposite effect: Local breakdown of the time-reversal symmetry in a superconducting state that preserves time-reversal symmetry \cite{Carlstrom2011length}.

In addition to the possible vanishing of the subdominant component, there is also a competition between the potential and kinetic phase-locking term that can restore the time-reversal symmetry in an external field. The potential phase-locking term reads as
\Equation{Eq:Potential-Phase-locking}{
	\frac{c_{12}}{2}\big(\psi_1^{*2}\psi_2^2+c.c.\big) = 
	c_{12} |\psi_1|^2|\psi_2|^2\cos2\varphi_{12}    \,,
}
and since the coefficient $c_{12}>0$, this promotes a relative phase $\varphi_{12}=\pm\pi/2$. The kinetic phase-locking term, on the other hand, is much less intuitive (see remark \footnote{ Note that the competition between kinetic and potential phase-locking terms can be made more explicit by eliminating the mixed-gradient term via another field redefinition, at the cost of a more complex potential, see details for example in \cite{Garaud2022effective}.}). It originates in the mixed gradient term that can be rewritten as
\Align{Eq:MixedGradient}{
	&\frac{k_{12}}{2}\big((\D\psi_1)^*\cdot\D\psi_2+c.c. \big) = \\
	&k_{12}\cos\varphi_{12}\Big[ \Grad|\psi_1|\cdot\Grad|\psi_2|+|\psi_1||\psi_2| \eth\varphi_1\cdot\eth\varphi_2\Big]   \nonumber \\
	-&k_{12}\sin\varphi_{12}\Big[ |\psi_2|\eth\varphi_2\cdot\Grad|\psi_1|
	-|\psi_1|\eth\varphi_1\cdot\Grad|\psi_2|    \Big]    \,,\nonumber
}
where $\eth \varphi_a=\Grad\varphi_a+e\A$. The kinetic phase-locking term couples the relative phase to density and phase variations and to the vector potential. The kinetic phase-locking term obviously plays no role in the ground state. However, in the presence of density or phase gradients, it will typically promote a relative phase of $\varphi_{12}=0,\pi$ and, thus, competes with the potential phase-locking term \Eqref{Eq:Potential-Phase-locking}. If the gradients are strong enough, then the kinetic phase-locking term \Eqref{Eq:MixedGradient} can dominate over the potential term and, thus, restore time-reversal symmetry. 

Depending on its stiffness, determined by the gradient coefficients, the subdominant component $\psi_1$ is suppressed in different manners. As illustrated in Figs.~\ref{Fig:Field-heated-averages} and \ref{Fig:Field-heated}, for the parameter set $\mbf{(a)}$, $\psi_1$ may survive, but the time-reversal symmetry is restored as a result of the kinetic phase-locking term. Alternatively, as illustrated in the case of the parameter set $\mbf{(b)}$, the subdominant component $\psi_1$ is gradually suppressed from the bulk, until fully vanishing. In both situations, this model is qualitatively consistent with the observed suppression of the spontaneous Nernst signal in magnetic field [Fig.~\ref{fig:rxx_Nernst_SH}(b)].

\begin{acknowledgments}
We would like to thank B. Mehlhorn for his contribution to the NMR measurements. 
The computations were enabled by resources provided by the National Academic Infrastructure for Supercomputing in Sweden (NAISS), partially funded by the Swedish Research Council through grant agreement no. NAISS 2023/3-11.
This work was partially performed at the Swiss Muon Source (S$\mu$S), PSI, Villigen. EB was supported by the Swedish Research Council Grants  2022-04763, by Olle Engkvists Stiftelse, and partially by the Wallenberg Initiative Materials Science
for Sustainability (WISE) funded by the Knut and Alice Wallenberg
Foundation. YW and VG are supported by the NSFC grants 12374139 and 12350610235.
We acknowledge support from the Deutsche Forschungsgemeinschaft (DFG)
through SFB 1143 (Project No.\ 247310070) and the
W\"{u}rzburg-Dresden Cluster of Excellence on Complexity and 
Topology in Quantum Matter--$ct.qmat$ (EXC 2147, Project No.\ 390858490),
as well as the support of the HLD at HZDR, member of the European
Magnetic Field Laboratory (EMFL). 

\end{acknowledgments}

\clearpage


%

\end{document}